\documentclass[
reprint,
superscriptaddress,
twocolumn,
 amsmath,amssymb,
 aps,
 pra
]{revtex4-2}
\usepackage{xcolor}
\usepackage{amsmath}

\usepackage{physics}
\usepackage{braket}
\usepackage{graphicx}
\usepackage{dcolumn}
\usepackage{bm}
\usepackage{placeins}
\usepackage{float}
\usepackage{amssymb}
\usepackage{amsthm}

\newtheorem{lemma}{Lemma}
\newtheorem{theorem}{Theorem}

\makeatletter
\@ifundefined{proof}{
  \newenvironment{proof}{\par\noindent\textit{Proof.}\ }{\hfill$\square$\par}
}{}
\makeatother
\makeatletter
\renewcommand*\env@matrix[1][*\c@MaxMatrixCols c]{
  \hskip -\arraycolsep
  \let\@ifnextchar\new@ifnextchar
  \array{#1}}
\makeatother
\DeclareMathOperator{\Erf}{Erf}
\DeclareMathOperator{\arctanh}{arctanh}
\usepackage{hyperref}
\usepackage{cleveref}

\begin{document}

\preprint{APS/123-QED}

\title{Fragility of Magic State Distillation under Imperfect Measurements}
\author{Yunzhe Zheng}
  \email{yunzhe.zheng@yale.edu}
 \affiliation{Department of Physics, Tsinghua University, Beijing, 100084, China}
  \affiliation{Department of Applied Physics, Yale University, New Haven, Connecticut, 06520, USA}
\author{Yuanchen Zhao}
\email{yc-zhao21@mails.tsinghua.edu.cn}
 \affiliation{Department of Physics, Tsinghua University, Beijing, 100084, China}
 \affiliation{Frontier Science Center for Quantum Information, Beijing 100184, China}
\author{Dong E. Liu}
 \email{dongeliu@mail.tsinghua.edu.cn}
 \affiliation{Department of Physics, Tsinghua University, Beijing, 100084, China}
 \affiliation{Frontier Science Center for Quantum Information, Beijing 100184, China}

\date{\today}

\begin{abstract}
\textbf{Abstract}\par
\medskip
Magic state distillation (MSD) is the leading approach for producing the non-Clifford resources required for universal fault-tolerant quantum computation. Most prior analyses of MSD assume ideal projective stabilizer measurements, but this assumption becomes questionable on near-term hardware, where measurement fidelity is limited and large quantum error-correcting codes are unavailable. We  consider imperfect measurements that are modelled as noisy projector controlled by \textit{measurement strength}, which can be regarded as inverse of measurement noise, and develop a general framework for analyzing MSD under imperfect stabilizer measurements. We focus on MSD protocols that are based on CSS codes with transversal non-Clifford gates, and show that MSD exhibits a measurement-strength threshold that is distinct from the previously known threshold on input-state error. When measurement strength is below a critical threshold, MSD loses its distillation power entirely; When measurement strength is above this threshold, distillation under imperfect measurements remains possible, but the asymptotic target states may deviate from the ideal magic states. In particular, this deviation is at most first-order biased. Moreover, we show that imperfect measurements with finite measurement strength reduces the distillation efficiency to linear, implying exponentially larger distillation overheads than in the ideal-measurement case. We further show that this fragility can be mitigated generally up to the code capacity of MSD protocols by choosing the stabilizer generators to measure in the \textit{standard form}. Our results reveal fundamental constraints imposed by imperfect measurements on MSD and provide guidance for designing more robust distillation protocols in realistic quantum hardware.

\end{abstract}

\maketitle

\section*{Introduction}

Magic state distillation (MSD) \cite{bravyiUniversalQuantumComputation2005, reichardtQuantumUniversalityMagic2005, reichardtQuantumUniversalityState2009, bravyiMagicstateDistillationLow2012} has been widely recognized as a central routine in fault-tolerant quantum computation, owing to the fundamental trade-off between universality and transversality \cite{zengTransversalityUniversalityAdditive2007, eastinRestrictionsTransversalEncoded2009}. Since Clifford gates are the most natural choice for transversal operations \cite{moussaTransversalCliffordGates2016, landahlFaulttolerantQuantumComputing2011a, paetznickFaulttolerantAncillaPreparation2013, krishnaFaulttolerantGatesHypergraph2021a}, non-stabilizer magic states are required to promote Clifford circuits to universal gate set at the logical level. Magic state distillation prepares high-fidelity magic states with arbitrarily low error by consuming many noisy copies, thereby enabling quantum advantage using fault-tolerant Clifford gates supplemented by noisy non-Clifford resources.

Magic state distillation relies critically on measurements and post-selection. During the distillation process, stabilizer measurements are performed on the noisy inputs, and only certain measurement outcomes herald successful distillation of higher-quality outputs. In this sense, magic state distillation shares the same spirit of quantum error detection, where measurement outcomes determine whether an error is detected or suppressed. Thus, measurement fidelity directly impacts distillation performance and the resource cost of preparing magic states. Most prior analyses have assumed ideal projective measurements \cite{bravyiUniversalQuantumComputation2005, bravyiMagicstateDistillationLow2012, hastingsDistillationSublogarithmicOverhead2018, haah2018codes, willsConstantOverheadMagicState2024}, justified by the belief that logical measurement errors can be suppressed with sufficiently large code distances in the long-term fault-tolerant regime. However, in realistic near-term hardware, measurement infidelity remains significant and qubit counts are limited to host high-distance error-correcting codes. In this setting, even logical measurements cannot be regarded as perfectly projective.

Imperfect measurements are a standard noise mechanism that have been studied extensively in open quantum systems
\cite{clerkIntroductionQuantumNoise2010, lloydQuantumFeedbackWeak2000,
shabaniContinuousMeasurementNonMarkovian2014}, quantum circuits
\cite{zhuNishimorisCatStable2023, PRXQuantum.4.030317,
zhao2023latticegaugetheorytopological, PhysRevB.110.064301}, and entanglement
distillation \cite{durQuantumRepeatersBased1999}. Their ubiquity is physically
natural, because measurements require coupling the system to an external probe, and the
resulting non-unitary interaction is rarely ideal in realistic devices.
However, their impact on MSD has not yet been characterized thoroughly. Some existing
studies analyzed MSD performance under noisy measurements and Clifford gates \cite{brooksQuantumErrorCorrection2013,
jonesMultilevelDistillationMagic2013a, jochymoconnorRobustnessMagicState2013,
litinskiMagicStateDistillation2019}, but these are mainly conducted through numerical studies of
specific protocols and these results may not extend to general MSD protocols. As recent experimental progress bringing MSD from far long-term fault-tolerant regime to practical near-term implementation \cite{brown_advances_2023,rodriguez_experimental_2024}, a general analytical framework is needed to understand the impact of imperfect measurements on MSD performance. 

In this work, we develop an analytical framework for studying the effect of imperfect measurements in MSD. We focus on a broad class of MSD protocols based on CSS codes with transversal non-Clifford gates, which includes the recent low-overhead MSD protocols \cite{bravyiMagicstateDistillationLow2012,hastingsDistillationSublogarithmicOverhead2018,willsConstantOverheadMagicState2024}, using the stabilizer reduction \cite{campbellStructureProtocolsMagic2009} description of MSD. we model imperfect stabilizer measurements as non-projective logical measurements controlled by a noise parameter we refer to as \textit{measurement strength}, which can be understood as the inverse of measurement noise. Our analysis then yields the following principal findings:

First, we find a critical threshold of measurement strength for MSD performance: When measurement strength is below the threshold, MSD cannot distill any input states to better output states and recursive distillation converges only to the maximally mixed state. Any MSD protocol fails entirely in this regime. When measurement strength is above the threshold, distillation under imperfect measurements remains possible for certain input states, but target state of MSD will be noisy. We also prove that the target states would only suffer at most first-order biased noise. 
Notably, the threshold for measurement strength is fundamentally distinct from the previously known threshold behavior in MSD \cite{bravyiUniversalQuantumComputation2005, reichardtImprovedMagicStates2005, campbell_magic-state_2012, meierMagicstateDistillationFourqubit2012}, where the notion of threshold has referred almost exclusively to the error of input states. Even when the input error is below the conventional threshold, imperfect measurements can destroy the convergence of the protocol. This distinction highlights the need for high-quality measurements to distill high-fidelity magic states in practice.

Second, we prove that any degree of measurement imperfection causes a severe degradation in distillation efficiency, which is defined to be the order of error suppression in MSD protocols, to linear order.  This degradation dramatically increases the distillation cost, which is defined as the average number of raw states needed to distill an $\epsilon$-good output state.  In the ideal measurement case, the cost scales polylogarithmically as ($O(\log^\gamma(1/\epsilon))$) for some constant $\gamma$, while in the noisy measurement case, the cost becomes a far less favorable polynomial one ($O((1/\epsilon)^\tau)$) for some constant $\tau$. As a result, the resource overhead for distilling magic states under imperfect measurements is exponentially higher than the ideal case.

Third, we propose a protocol-agnostic strategy to mitigate the impact of imperfect measurements in MSD. By selecting a specific generator set for the code stabilizers to measure, known as the \textit{standard form} \cite{gottesmanStabilizerCodesQuantum1997,nielsenQuantumComputationQuantum2010}, the robustness to measurement noise is maximized up to the code capacity of the distillation protocols. In principle, this strategy can be implemented without additional ancilla qubits or gate overhead, but only by modifying the compilation of the distillation circuit.

We remark that these results distinguish our work from prior studies of noisy Clifford operations and measurement faults in MSD \cite{brooksQuantumErrorCorrection2013,jonesMultilevelDistillationMagic2013a,jochymoconnorRobustnessMagicState2013,litinskiMagicStateDistillation2019}. Conceptually, previous work mainly focused on specific distillation protocols under concrete circuit-level implementations and tracked the output error after a finite number of distillation rounds. By contrast, we study a broad class of MSD protocols based on CSS codes with transversal non-Clifford gates and model imperfect stabilizer measurements directly as non-projective measurements. This perspective lets us characterize asymptotic target states, measurement-strength thresholds, and distillation efficiency independently of a particular circuit realization. Technically, while much of the prior literature relies primarily on circuit-level numerical simulation, our main conclusions are analytical, stated as theorems with formal proofs, and are supported by numerical simulations based on the dynamical-system representation of MSD rather than on a specific circuit-level noise model. We believe our results can help understand robustness of known MSD protocols and design new MSD protocols suitable for realistic quantum hardware.

\section*{Results}

\begin{figure}[t]
    \centering
    \includegraphics[width=.85\linewidth]{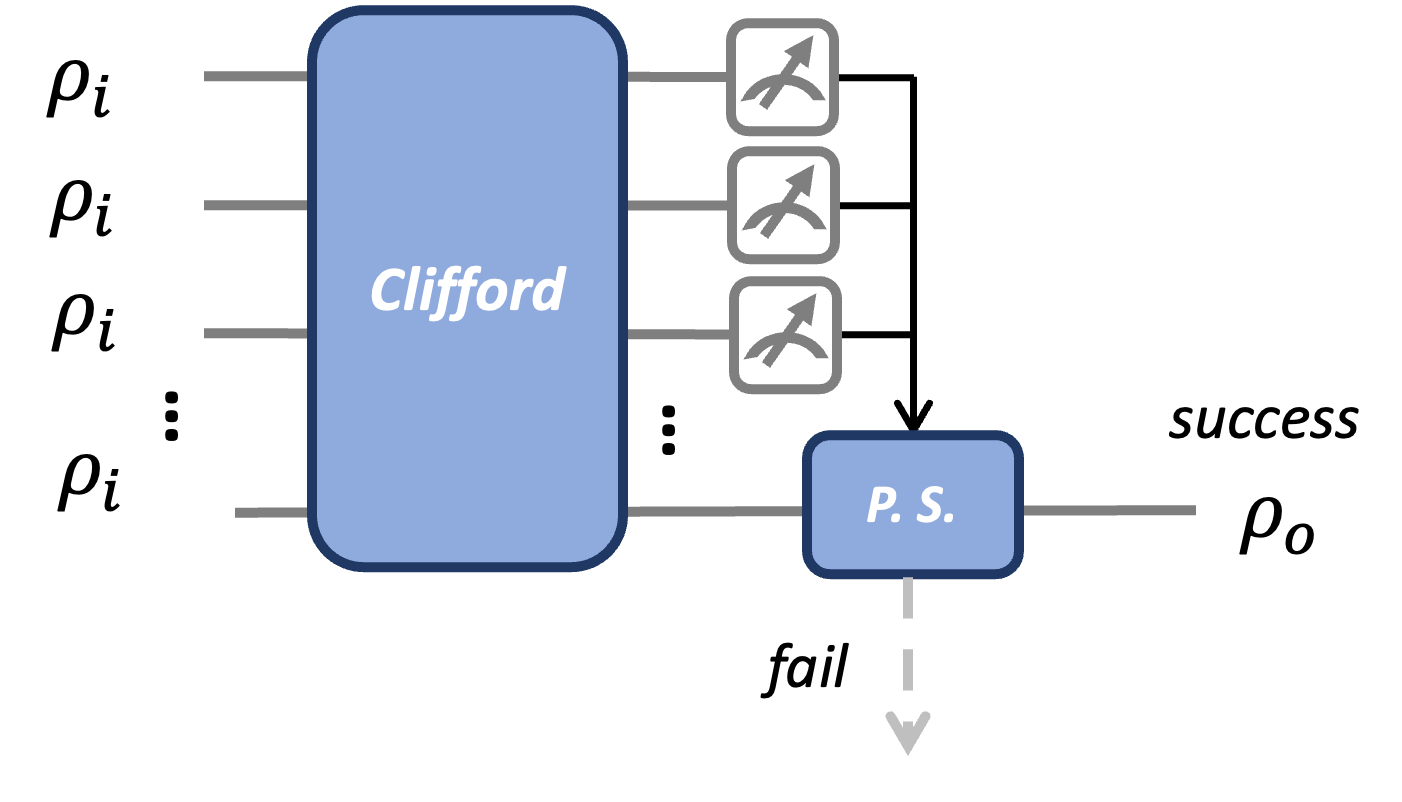}
    \caption{\textbf{Stabilizer reduction scheme.} We consider the input to be multiple copies of noisy magic states, and a Clifford circuit associated with the protocol code $\mathcal{Q}$ is applied on the input state followed by single Pauli measurements on all ancillary qubits. Post-selecting on all $+1$ outcome heralds successful output state.}
    \label{fig:stabilizer-reduction}
\end{figure}

\subsection*{Theoretical formalism}

All known MSD protocols can be classified as \textit{Stabilizer Reduction} (SR) protocols \cite{campbellStructureProtocolsMagic2009} described by stabilizer codes. For an $[[n, k]]$ stabilizer code $\mathcal{Q}$ with stabilizer generator $\{g_i\}$, it describes the following $n$-to-$k$ SR protocol (distill an $n$-qubit input state into a $k$-qubit state): 
\begin{itemize}
    \item Prepare an $n$-qubit input tensor product state $\rho_{\mathrm{in}}=\rho_{\mathrm{i}}^{\otimes n}$. $\rho_{\mathrm{i}}$ is the noisy magic states and can be, for instance, $\ket{T}=T\ket{+}$ states under Pauli noise.
    \item Measure every stabilizer generator $g_i$ for $i=1,2...,(n-k)$ on $\rho_{\mathrm{in}}$.
    \item Post-select on all $+1$ outcomes.
    \item Decode the post-selected state with a Clifford decoding circuit.
\end{itemize}
Although some measurement outcomes other than all $+1$ may also herald successful output states up to known Clifford gauge corrections \cite{bravyiUniversalQuantumComputation2005}, throughout this work we only consider the all-$+1$ without loss of generality as they don't affect the distillation performance except for the post-selection rate. 

From the mathematical perspective, measuring every stabilizer and post-selecting on all $+1$ outcome would project the input states on to the codespace of $\mathcal{Q}$. The successfully post-selected state is therefore 
\begin{equation}
    \rho_{\mathrm{p}} = P_\mathcal{Q}\rho_{\mathrm{in}}P_\mathcal{Q}/\Tr[P_\mathcal{Q}\rho_{\mathrm{in}}],
\end{equation}
where the codespace projector $P_\mathcal{Q}$ is given by
\begin{equation}
    P_\mathcal{Q} = \prod^{\bar{n}}_{i=1}\frac{I+g_i}{2} = \frac{1}{2^{\bar n}}\sum^{2^{\bar n}}_{j=1} s_j,
    \label{eq:projector}
\end{equation}
where $\bar{n}=n-k$ and $s_j$ are all the elements in the stabilizer group defined on $\mathcal{Q}$. 
Finally, as the post-selected state $\rho_{\mathrm{p}}$ is an $n$-qubit state that encode $k$-qubit information, we should decode it for output usage. For example, if the output state is supposed to be a single-qubit state, i.e. $k=1$, the final successfully distilled state from $\rho_{\mathrm{in}}$ takes the form  
\begin{equation}
 \rho_{\mathrm{o}} = \frac{1}{2}(I + \Tr[\rho_{\mathrm{p}}\bar{X}]X+\Tr[\rho_{\mathrm{p}}\bar{Y}]Y+\Tr[\rho_{\mathrm{p}}\bar{Z}]Z)   
\end{equation}
where $\bar{X},\bar{Y},\bar{Z}$ are the logical Pauli operators for code $\mathcal{Q}$ and $X,Y,Z$ are single-qubit Pauli matrices. The expression for multi-output $(k>1)$ can be similarly obtained by enumerate each logical qubit and corresponding logical operators. 

\begin{figure*}[t!]
    \centering
    \includegraphics[width=.98\textwidth]{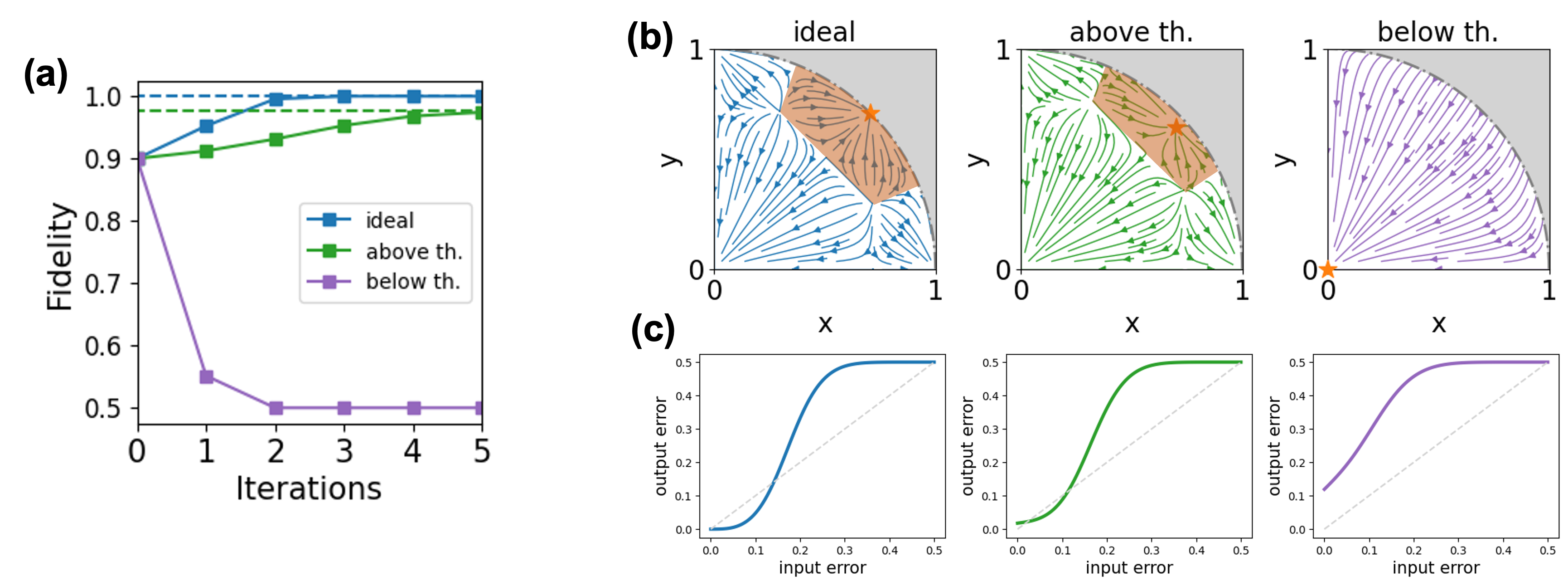}
    \caption{\textbf{Critical distillation behavior of the $[[15, 1, 3]]$ protocol for different measurement strength.} (a) Iterative distillation behavior for the $[[15, 1, 3]]$ protocol. When measurement strength of the noisy measurement is above the critical threshold, distillation is still achievable but with a fidelity cap (dashed line). Distillation fails entirely if measurement strength is below the critical threshold. The initial state is set to be depolarized $\ket{T}$ state with error rate 0.1. (b) Flow diagram of the $[[15, 1, 3]]$ MSD protocol on the $x$-$y$ cross section of the Bloch sphere. The orange stars denote the target states to distill into and the orange area denotes the convergence region for the target states. $\beta=2$($\beta=1$) is used for the above(below) threshold simulation. The threshold is $\beta^*\approx 1.74$. (c) Input-output error relation for the $\ket{T}$ state under depolarizing noise.}
    \label{fig:critical-distillation}
\end{figure*}

Technically, the state $\rho_{\mathrm{o}}$ can be obtained by applying a Clifford decoding circuit associated with $\mathcal{Q}$ to $\rho_{\mathrm{p}}$ and tracing out all ancilla qubits \cite{zhengMagicStateDistillation2024}, and the decoding circuit can be explicitly constructed from the stabilizer of $\mathcal{Q}$ using Gottesman's algorithm \cite{gottesmanStabilizerCodesQuantum1997}. Besides, as stabilizer measurement can propagate through the decoding circuit to single Pauli $Z$s, we can postpone the stabilizer measurement after the decoding circuit and perform Pauli $Z$ measurements at the end of the circuit. Therefore, the practical scheme for stabilizer reduction protocols is shown in Figure 1. Notably, encoding the input state to logical states of $\mathcal{Q}$ is not necessary for SR protocols; see Supplementary Note~\ref{app:encoding}. An important class of SR protocols is CSS codes that admit transversal non-Clifford diagonal gates, such as the $T$ gate or the $CCZ$ gate. These CSS codes allow distillation of the corresponding magic states (e.g.  $\ket{T} = (\ket{0}+e^{i\pi/4}\ket{1})/\sqrt{2}$ or $\ket{CCZ} = CCZ\ket{+}^{\otimes 3}$), and all known practical or low-overhead MSD protocols fall into this class \cite{bravyiMagicstateDistillationLow2012, krishnaLowOverheadMagic2019, hastingsDistillationSublogarithmicOverhead2018, willsConstantOverheadMagicState2024}. 

In the ideal case, measurements are all projective and the projective operators for stabilizer generator $g$ are
\begin{equation}
    M_\pm(g) =\frac{I\pm g}{2}.
\end{equation}
The codespace projector $P_\mathcal{Q}$ can be constructed by the product of measurement operators for all generators $M_+(g_i)$, and codespace projection can be achieved by measuring every generator $g_i$ and post-selection on all +1. However, this might no longer hold for the noisy case. We consider the imperfect measurement model for Pauli operator $g$ to be :
\begin{equation}
\begin{split}
     \tilde{M}_\pm(g) &= f(\beta) M_\pm(g) + \sqrt{1-f(\beta)^2}M_\mp(g)   \\
     &\propto I \pm h(\beta) g
\end{split}
\end{equation}
where $f(\beta)$ and $h(\beta)$ are both noise coefficients and controlled by measurement strength $\beta$, which is defined to be the inverse of noise strength. The ideal case is recovered when $f(\beta)=1$ and $h(\beta)=1$, and the noisiest case is when $f(\beta)=\frac{1}{\sqrt{2}}$ and $h(\beta) = 0$. For example, the typical Gaussian-type noisy measurements \cite{zhuNishimorisCatStable2023} are described by $\tilde M_\pm(g, \beta) = \frac{\exp(\pm\beta  g/2)}{\sqrt{2\cosh \beta}}$, with $h(\beta) = \tanh(\beta/2)$ (also see Supplementary Note~\ref{app:gaussian}). Notably, the noisy measurement operators are no longer projectors, as $\tilde{M}_\pm^2 \neq \tilde{M}_\pm$ in the generic case.

The noisy codespace measurement operator for stabilizer code $\mathcal{Q}$ now becomes the product of all positive noisy measurement operators:
\begin{equation}
    \tilde M_{\mathcal{Q}} = \prod^{\bar{n}}_{i=1}\tilde{M}_+(g_i) \propto \prod^{\bar{n}}_{i=1}(I + h(\beta) g_i).
    \label{eq: weak_meas_op}
\end{equation}
The noisy post-selected states are now
\begin{equation}
    \tilde\rho_{\mathrm{p}} \propto \tilde{M}_\mathcal{Q}\rho_{\mathrm{in}}\tilde{M}_\mathcal{Q}/\Tr[\tilde{M}^2_\mathcal{Q}\rho_{\mathrm{in}}],  
    \label{eq: noisy_rho_p}
\end{equation}
 and the decoded state for output will then be given by 
\begin{equation}
 \tilde \rho_{\mathrm{o}} = \frac{1}{2}(I + \Tr[\tilde\rho_{\mathrm{p}}\bar{X}]X+\Tr[\tilde\rho_{\mathrm{p}}\bar{Y}]Y+\Tr[\tilde\rho_{\mathrm{p}}\bar{Z}]Z).
\end{equation}
It's noted the coefficient terms in $\tilde \rho_{\mathrm{o}}$  only depend explicitly on $\Tilde{M}^2_\mathcal{Q}$. Therefore, between the noiseless and noisy stabilizer reduction, the only difference is to replace the codespace projector $P_\mathcal{Q}$ with the noisy "projector" counterpart $\Tilde{M}^2_\mathcal{Q}$ (instead of $\tilde M_\mathcal{Q}$). Using the definition from Eq. \eqref{eq: weak_meas_op}, we have

 \begin{equation}
     \tilde{M}_\mathcal{Q}^2 \propto  \prod^{\bar{n}}_{i=1}(I + h^2(\beta) g_i) = \sum^{2^{\bar n}}_{j=1} (h^2(\beta))^{\gamma_j} s_j,
      \label{eq:m_square}
 \end{equation} 
 where $s_j$ are all stabilizer elements and $\gamma_j$ is the number of generators required to generate stabilizer $s_j$. In contrast to the ideal codespace projector in Eq. \eqref{eq:projector}, the coefficients of terms in the noisy "projector" in Eq. \eqref{eq:m_square} are no longer uniform. Instead, the coefficients depend on the specific stabilizer generators being measured. Each stabilizer term acquires a prefactor $h^2(\beta)$ that is smaller than 1, determined by the number of generators needed to construct it. For the trivial stabilizer $I$, the coefficient is still 1. For generators themselves as stabilizer elements, the coefficient is $h^2(\beta)$. Therefore, all stabilizer elements in the noisy "projector" suffers from a non-uniform squeeze of coefficients. As projection is essential for MSD process, it also indicates that the noisy MSD process can depend sensitively on the choice of stabilizer generators to measure.

\subsection*{Threshold for measurement noise}

Our theoretical derivation allows us to simulate MSD protocols under imperfect measurements by mapping to dynamical systems, which allows us to visualize the MSD protocols using flow diagram. The mapping method was initially proposed for the ideal measurement case \cite{zhengMagicStateDistillation2024}, but here we integrate our theoretical framework and show its applicability for the noisy case (see Methods). The non-trivial fixed points in the mapped dynamical systems correspond to the \textit{target states} in the MSD protocols, i.e. the best output state after infinite times of iteration, and the convergence rate for the non-trivial fixed points stands for the \textit{distillation efficiency} (error suppression rate) of the MSD protocols.

Firstly, we consider the $[[15, 1, 3]]$  protocol with canonical generator description that can be understood as a 3D color code \cite{kubicaUniversalTransversalGates2015}. The detailed code description can also be found in Supplementary Note~\ref{app: code_description}. Again, we emphasize that the choice of stabilizer generators can impact the distillation performance, and we specify our generator choice for completeness. We use the Gaussian-type noise model for simulation, and $\beta\in[0,\infty)$ is the measurement strength. Ideal case is recovered when $\beta\rightarrow\infty$, and the first-order noise strength scales with $e^{-2\beta}$ as $\tilde{M}_+(g)^2 \propto M_+(g) + e^{-2\beta} M_-(g)$. In Figure 2(a), We simulated the iterative distillation behavior under different measurement strength. We observe a critical behavior for the measurement strength: When measurement strength is finite but above a threshold, input state can be distilled into better states capped by a finite infidelity. When measurement strength is below a threshold, distillation always fails and no input states can be distilled into better states. The numerical value for critical strength is $\beta^* \approx 1.74$.

We then plot the flow diagram for the $[[15, 1, 3]]$ codes in the $xy$ cross-section of the Bloch sphere in Figure 2(b). We find out the target state will move toward inside the Bloch sphere and the convergence region for the target state will shrink when measurement strength (noise) is reduced (increased). Furthermore, we find an abrupt jump of the target state to the zero point at the critical strength $\beta_*$. We also observed similar behavior for the $[[14, 2, 2]]$ protocol \cite{bravyiMagicstateDistillationLow2012}, and the critical strength is $\beta^* \approx 1.89$. In Figure 2(c), we plot the relation between output error and the input error under depolarizing noise for reference on the distillable regime of input states.

We claim the critical threshold behavior for measurement strength must be a ubiquitous phenomenon for MSD protocols. For all known MSD protocols, there exists a finite fidelity threshold \cite{bravyiUniversalQuantumComputation2005, reichardtImprovedMagicStates2005, meierMagicstateDistillationFourqubit2012, campbell_magic-state_2012} only above which the input states can distill into better output states. Therefore, the convergence region to the target state is finite-size as well. In the presence of measurement noise, the fidelity threshold will increase and the convergence region will shrink with decreased measurement strength, while the target state will become more noisy and move inside as we see in Figure 2(b). The critical point occurs when the inner boundary of the convergence region touches the target state, which corresponds to the threshold for measurement strength. 

\subsection*{Fidelity cap of target states}

\begin{figure}[th]
    \centering
    \includegraphics[width=\linewidth]{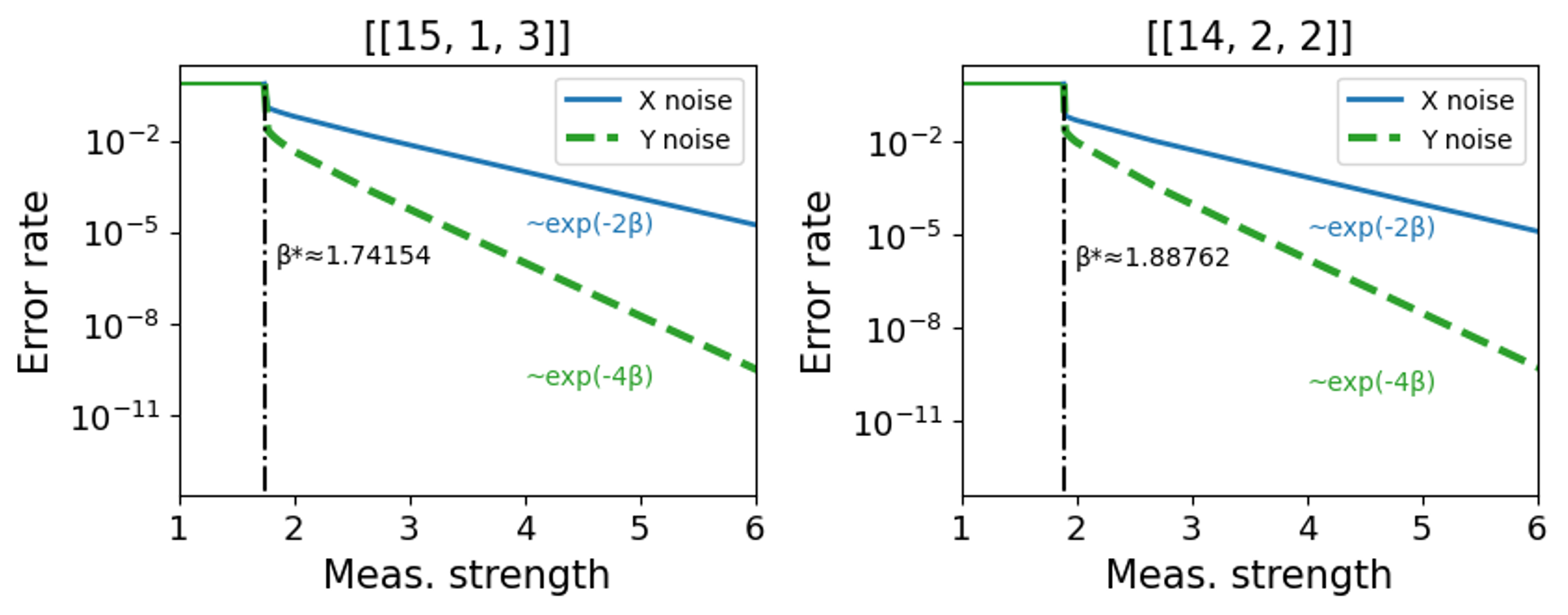}
    \caption{\textbf{Deviation of target states.} For both $[[15, 1, 3]]$ and $[[14, 2, 2]]$ protocols, the target states deviates from the ideal $\ket{T}$ states under imperfect measurements. The deviation is also dominated by the $X$ noise as the $Y$ noise is significantly in the smaller order. $\exp(-2\beta)$ is the factor for first-order noise in the simulated noise model.}
    \label{fig:target-state-deviation}
\end{figure}

As we see in Figure 2(b), the target states under imperfect measurements deviates from the ideal state but not in a symmetric way regarding the $x$ and $y$ axis. Now we show that the first-order deviation must be biased for transversal MSD protocols based on CSS codes. We call a MSD protocol \textit{transversal} if the associated stabilizer code has transversal implementation non-Clifford gate and it distill into magic states corresponding to the non-Clifford gate. For example, both $[[15, 1, 3]]$ \cite{bravyiUniversalQuantumComputation2005} and $[[14, 2, 2]]$ \cite{bravyiMagicstateDistillationLow2012} protocol are transversal for the $\ket{T}$ state.
Notably, all known low-overhead MSD protocols \cite{hastingsDistillationSublogarithmicOverhead2018, willsConstantOverheadMagicState2024} are transversal protocols.

\begin{theorem}[\textbf{Biased target state}]
    For transversal MSD protocols based on CSS codes, the impact of first-order measurement noise, if presented, must be biased on the output states.
    \label{th: bias}
\end{theorem}
\begin{proof}[Proof outline]
From Eq. \eqref{eq: weak_meas_op}, $\tilde M_{+}(g,\beta)$ can be written as a linear combination of $M_+(g)$ and $M_-(g)$. As measurements are independent, first-order analysis only considers when one of the measurements is erroneous and the measurement operator $M_+(g)$ is replaced by $M_-(g)$.
Therefore, the post-measurement state $\rho_{\mathrm{p}}$ in Eq. \eqref{eq: noisy_rho_p} can be written as a linear combination of $\bar{P}\rho_{\mathrm{in}}\bar{P}$ and $\bar{P}_{g}\rho_{\mathrm{in}}\bar{P}_{g}$ and their non-diagonal terms, where $\bar{P}$ is the codespace projector and $\bar{P}_{g}$ is the subspace projector stabilized by every stabilizer generator but anti-stabilized by the corrupted generator $g$. Consider $\rho_{\mathrm{in}}=(\ket{\theta}\bra{\theta})^{\otimes n}$ to be the ideal input state, where $\ket{\theta}=R_{\mathrm{Z}}(\theta)\ket{+}$ and $R_{\mathrm{Z}}(\theta)=\exp(-i\theta Z/2)$ is transversal for the CSS codes.  If the corrupted generator is $X$-type, i.e., $g = g_{\mathrm{x}}$ , we can prove
\begin{equation}
\bar{P}_{g_{\mathrm{x}}}\ket{\theta}^{\otimes n}   = \Tilde{Z}   R_{\mathrm{Z}}(\theta)^{\otimes n}\bar{P}\Tilde{Z}\ket{+}^{\otimes n} = 0,
\label{eq: X_type}
\end{equation}
where we used the fact that logical operations commute with codespace projector (Supplementary Note~\ref{app:lemmas}) and $\tilde{Z}$ is a $Z$-type subspace-transfer operator such that  $\bar{P}=\tilde Z\bar{P}_{g_{\mathrm{x}}} \tilde Z$ (Supplementary Note~\ref{app:lemmas}). As $\ket{+}^{\otimes n}$ is stabilized by all $X$-type stabilizers, $\Tilde{Z}\ket{+}^{\otimes n}$ must be anti-stabilized by at least one $X$-type stabilizers and therefore $\bar{P}\Tilde{Z}\ket{+}^{\otimes n}=0$.
If the corrupted generator is $Z$-type instead, i.e.,  $g = g_{\mathrm{z}}$,  we show
\begin{equation}
\bar{P}_{g_{\mathrm{z}}}\ket{\theta}^{\otimes n}   = \Tilde{X} \bar{P} \Tilde{X} R_{\mathrm{Z}}(\theta)^{\otimes n}\ket{+}^{\otimes n} \propto \Tilde{X}\bar{P}\ket{\theta}^{\otimes n} \label{eq: Z_type0}
\end{equation}
where $\tilde{X}$ is a $X$-type subspace-transfer operator such that $\bar{P}= \tilde{X}\bar{P}_{g_{\mathrm{z}}} \tilde{X}$. As $\tilde{X}$ always commutes with logical $X$ operators, its effect on the decoded states must either be a logical $X
$ or identity operator. Therefore, there is no first-order $Z$-type noise on the output state. Full proof can be found in Supplementary Note~\ref{app:proofs}.
\end{proof}

We then simulate the deviation of the target states for both the $[[15, 1, 3]]$ and $[[14, 2, 2]]$ protocol. Because the target states are $\ket{T}$ states and $\braket{Z}_{\ket{T}}=0$, we can decompose the noise on $\ket{T}$ states into only Pauli $X$ and $Y$ noise. As shown in Figure 3, the main contribution to target state deviation is from Pauli $X$ noise. $Y$ noise serves as the smaller order. Therefore, it corroborates with the Theorem \ref{th: bias} and reveals the intrinsic biased noise structure in MSD under imperfect measurements.

It's known that the stabilizer codes used for MSD often have asymmetric code distance for $X$ and $Z$ errors, and logical biased noise could be presented due to the asymmetric distance \cite{fazio_logical_2024, ruiz_unfolded_2025, lee2024low}. However, we remark that the first-order bias in Theorem \ref{th: bias} doesn't come from this distance asymmetry. Instead, it is because only erroneous $Z$-type measurements would cause noisy output states, and erroneous $X$-type measurements are detectable and doesn't impact the output states.

\subsection*{Distillation efficiency and cost}

Next, we consider impact of imperfect measurement on the distillation efficiency. we simulate the convergence of recursive distilled states in Figure 4, and the convergence rate in the dynamical systems can be understood as the distillation efficiency in the distillation protocols. Surprisingly, we found out the distillation efficiency for both protocols degrades to linear under finite measurement strength. In another word, the error suppression relation from input error $\epsilon_{\mathrm{i}}$ to output error $\epsilon_{\mathrm{o}}$ will change from $\epsilon_{\mathrm{o}} = O(\epsilon^d_i)$ in the ideal case to $\epsilon_{\mathrm{o}} = k' \epsilon_{\mathrm{i}}$ with some prefactor $k'<1$. Notably, the canonical distillation cost scaling $O(\log^\gamma(1/\epsilon))$ only holds when the efficiency $d$ is at least quadratic ($d>1$). Linear efficiency would cause distillation cost to be $O((1/\epsilon)^\tau)$, which is exponentially higher (Figure 4(b,d)). The exponent is given by $\tau = \log(n/k)/\log(1/k')$ (see Methods).

Actually, the efficiency degradation turns to be a ubiquitous phenomenon for MSD protocols under imperfect measurements:
\begin{theorem}[\textbf{Linear efficiency}]
 For transversal MSD protocols based on CSS codes, the distillation efficiency will be linear under imperfect measurements.    
\label{th: linear}
\end{theorem}
We leave the full proof in Supplementary Note~\ref{app:proofs}. One important implication of the Theorem \ref{th: linear} is: The distillation cost for magic states might not preserve the good cost scaling if the ideal measurement assumption no longer holds. The overhead for distillation is quantified with the parameter $\gamma$ in the scaling $O(\log^\gamma(1/\epsilon))$. However, distillation efficiency is degraded to linear as long as the measurement is not perfectly projective, and the distillation cost turns to be $O((1/\epsilon)^\tau)$. This implies the cost for distilling magic states could be much higher in practice in contrast of the previous logarithmic understanding.

\begin{figure}[t!]
    \centering
    \includegraphics[width=.98\linewidth]{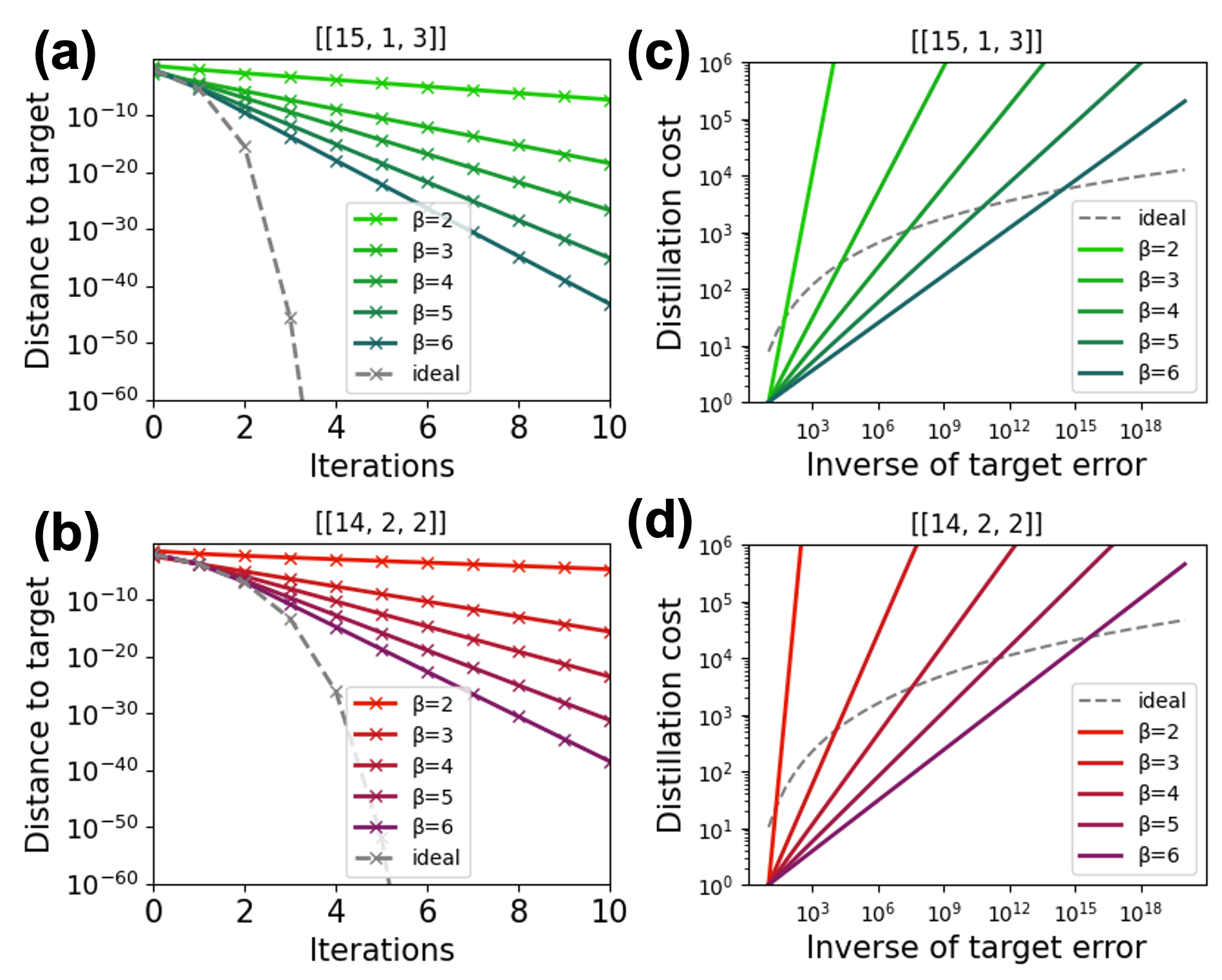}
    \caption{\textbf{Convergence and distillation cost.} (a-b) Convergence of distilled states toward the target states for $[[15, 1, 3]]$ and $[[14, 2, 2]]$ protocols. The convergence under ideal measurements is exponentially faster than the imperfect cases. (c-d) Distillation cost under imperfect measurements. As imperfect measurements cause linear efficiency, the distillation cost is also exponentially higher than the ideal case.}
    \label{fig:convergence-cost}
\end{figure}

\subsection*{Noise-resilient MSD}

We now provide a method to render MSD protocols more robust to measurement errors. Our method relies on measuring stabilizer generators in standard form, which is a specific way to describe stabilizer code using sympletic matrix. In particular, any $[[n, k]]$ CSS stabilizer codes can be described by the $\bar n \times 2n$ standard-form matrix \cite{nielsenQuantumComputationQuantum2010} as
\begin{equation}
    [g_1, g_2, ... ,g_{\bar{n}}]^T =  [H_{\mathrm{X}} | H_{\mathrm{Z}}] = \begin{bmatrix}[ccc|ccc]
        I & A_1 & A_2 & 0 & 0 & 0 \\
        0 & 0 & 0 & D & I & E
    \end{bmatrix}.
\end{equation}
The row of above matrix is $r$ (the rank of $H_{\mathrm{X}}$) and $\bar{n}-r$, and the column is divided by $r$, $n-k-r$ and $k$. $I$ is the identity matrix and $A_{1,2}, D, E$ are all non-zero matrices. The standard-form description can be efficiently obtained by Gaussian elimination and permutation from an arbitrary generator description using symplectic matrix.
\begin{theorem}[\textbf{Robust distillation}]
     For a distance-$d$ transversal MSD protocol based on CSS codes, it can be made robust to measurement errors up to order $d-1$ by measuring generators in the standard form.
    \label{th: standard}
\end{theorem}

We first explain the intuition for Theorem~\ref{th: standard} and then provide the proof outline.
In the proof outline of Theorem~\ref{th: bias}, we see imperfect measurements affect the output states through the corresponding subspace-transfer operators. If a subspace-transfer operator commutes with every logical Pauli operator, then it has trivial logical action and does not affect the output states. If it anti-commutes with at least one logical Pauli operator, then it acts nontrivially on the logical subspace and induces a Pauli error on the output state. Therefore, to minimize the impact of higher-order measurement errors, we want to choose a set of stabilizer generators to measure such that the subspace-transfer operators induced by all lower-weight measurement-error patterns still commute with the logical Pauli operators. Measuring generators in standard form achieves this because the subspace-transfer operators associated with the individual generators can be chosen to be weight-one, supported on distinct qubits, and commute with every logical operator (Supplementary Note~\ref{app:lemmas}). Hence any product of fewer than $d$ such operators remains commutative with every logical operator, implying that the protocols can tolerate up to order-$d-1$ measurement errors.

\begin{proof}[Proof outline]
Similar to the proof for Theorem \ref{th: bias}, we can assume there are $d'<d$ generators $\{g_{i_1}, g_{i_2},...,g_{i_{d'}}\}$ that get corrupted and the whole state is projected onto the corresponding subspace $\mathcal{Q}'$. Say $C_{i_j}$ is the subspace-transfer operator for $\bar {P}_{g_{i_j}}$, the subspace-transfer operator for $\mathcal{Q}'$ is then given by $\bar{C} = \prod^{d'}_{j=1} C_{i_j}$ and the projector for $\mathcal{Q}'$ is defined via $P_{\mathcal{Q}'} = \bar{C} \bar{P} \bar{C}.$ Without loss of generality we assume the first $m$ generators are $Z$-type and the remaining $d'-m$ generators are $X$-type. Define $R_{\mathrm{Z},j}(\theta)$ to be the $R_{\mathrm{Z}}(\theta)$ gate on qubit $j$, we have
\begin{equation}
\begin{split}
    P_{\mathcal{Q}'}\ket{\theta}^{\otimes n} &= \bar{C} \bar{P} \bar{C} R_{\mathrm{Z}}(\theta)^{\otimes n} \ket{+}^{\otimes n}\\
    & = \bar{C}   R_{\mathrm{Z}}(\theta)^{\otimes n} \bar{P}(\bigotimes^m_{j=1} R_{\mathrm{Z}, j}(-2\theta) )\bar{C} \ket{+}^{\otimes n}\\
    & \propto \bar{C}  R_{\mathrm{Z}}(\theta)^{\otimes n} \bar{P} \ket{+}^{\otimes n} = \bar{C}   \bar{P} \ket{\theta}^{\otimes n}.\\
\end{split}
\end{equation}
where we used the fact that all weight-$d'$ Z errors are detectable by $\bar{P}$ from the second row to the last row. As all $C_i$ are weight-one and commute with the standard-form logical operators based on Supplementary Note~\ref{app:lemmas}, $\bar{C}$ also commutes with the standard-form logical operators. Therefore, it will act as if logical identity after decoding and won't affect the output state. The MSD protocols can therefore tolerate imperfect measurement up to order $d-1$. Full proof is shown in Supplementary Note~\ref{app:proofs}.
\end{proof}

\begin{figure}[th]
    \centering
    \includegraphics[width=\linewidth]{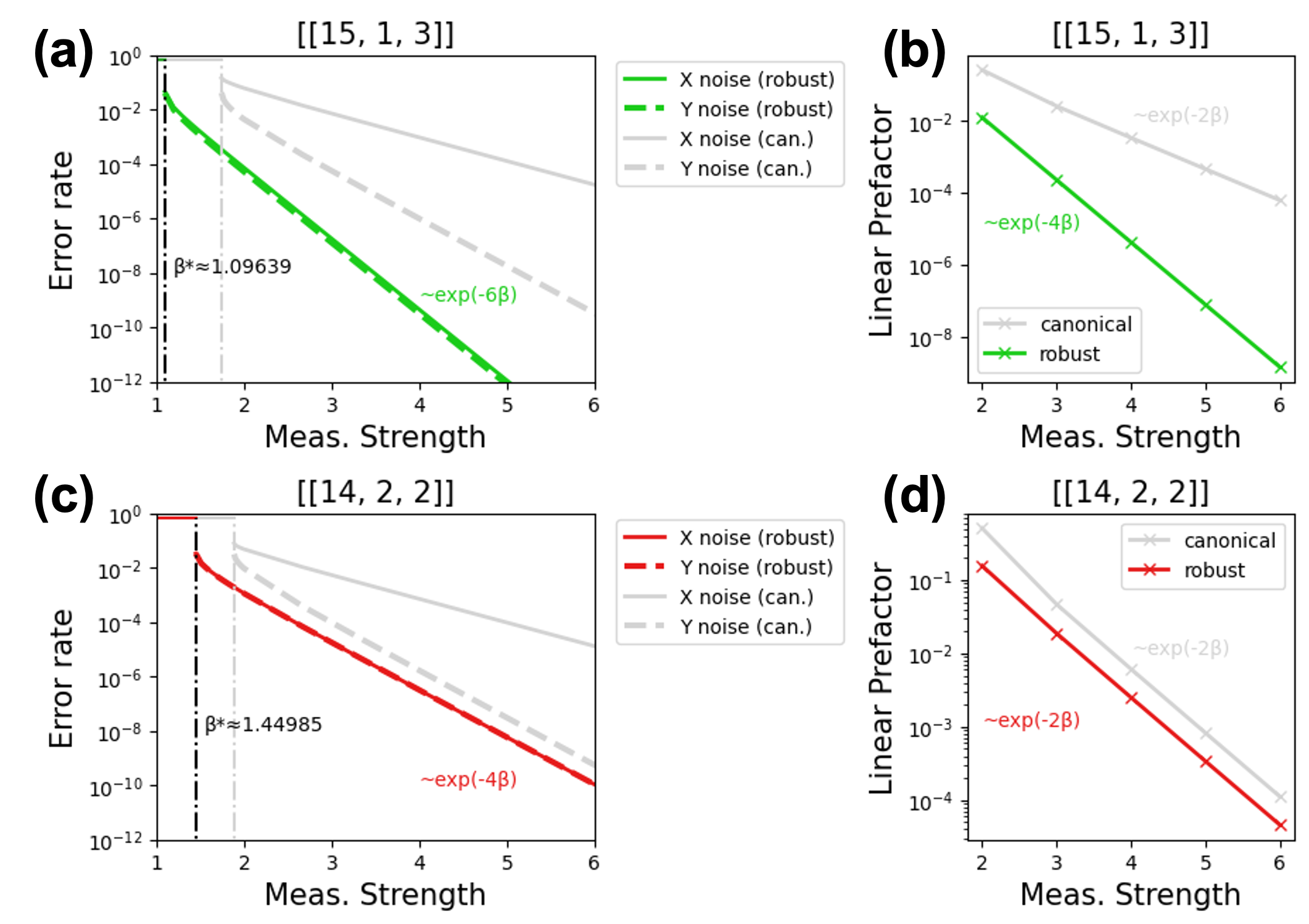}
    \caption{\textbf{Noise-resilient MSD protocols under imperfect measurements by measuring standard-form generators.} (a,c) Displacement between the noisy target states and the ideal state. (b,d) Scaling of the linear convergence prefactor. The grey line in all figures denotes the performance when measuring canonical generators.}
    \label{fig:standard-form-performance}
\end{figure}

We simulate both $[[15, 1, 3]]$ and $[[14, 2, 2]]$ MSD protocols with generators in standard form as shown in Figure 5 and compare the performance with the canonical generator choices (grey line). For both protocols, we gain better (lower) threshold for measurement strength, which means the protocols are able to tolerant more errors before they fail. Besides, the target states also suffer much lower deviation from the ideal state. As the first-order noise scales with $e^{-2\beta}$, both protocols saturate their code capacity to correct the noise on the target states. For the $[[15, 1, 3]]$ protocol, the deviation scales with $e^{-6\beta}$ as the distance is three, while the for $[[14, 2, 2]]$ protocol, the deviation scales with $e^{-4\beta}$ as the distance is two. Lastly, the linear convergence is also faster than the canonical case as shown in Figure 5(b,d). 

Remarkably, choosing standard-form generators to measure won't essentially add any substantial practical overhead. To measure standard-form generators for MSD, we only need to modify the Clifford circuits in Figure 1. The circuit difference between standard-form and canonical generators is several CX gates. This is because the standard form can be obtained from any generator set using Gaussian elimination: The subtraction in matrix row stands for a CX gate in the Clifford circuit, and the column swap stands for a SWAP gate that can be performed virtually. Even though the standard-form circuit might not be the one with minimal two-qubit gates depending on the supported native gates and compilation, the standard-form encoding circuit for an $n$-qubit code only uses $O(n^2/\log n)$ two-qubit gates and is optimal in the asymptotic case \cite{aaronson_improved_2004}.

\section*{Discussion}

\begin{table*}[t]
\centering
\begin{tabular}{c | c | c | c}
\hline
\textbf{Measurement strength} & \textbf{Target states} & \textbf{Distillation efficiency} & \textbf{Distillation cost} \\ \hline
\textit{infinite (ideal)} & pure magic states & code distance & $O(\log^{\gamma}(\epsilon^{-1}))$ \\ \hline
\textit{finite but above threshold} & noisy magic states & linear  & $O((\epsilon_{\mathrm{r}}/ \epsilon)^\tau)$ \\ \hline
\textit{below threshold} & fully mixed states & - & $\infty$ \\ \hline
\end{tabular}
\caption{Comparison of MSD performance. Distillation cost refers to the average number of raw magic states with error $\epsilon_{\mathrm{r}}$ needed to distill an $\epsilon$-good magic state.}
\label{tab:example}
\end{table*}

In this work, we established the framework for analyzing MSD under imperfect measurement and analyzed its performance both theoretically and numerically. We summarize the performance of MSD in different cases in Table \ref{tab:example}. We see that the presence of measurement noise will degrade the distillation efficiency to linear, which could ultimately lead to higher overhead of magic states from distillation. Besides, it also posts another perspective on the practicality of low-overhead MSD. It's noted the error suppression prefactor could be related with the measurement strength and the code parameter the protocols. Therefore, it would be interesting to investigate if the low-overhead MSD protocols with small $\gamma$ remains comparatively low-overhead for the parameter $\tau$ under the same measurement strength.

We remark that this work mainly focuses on logical-level MSD, where the measurement noise should be understood as logical noise remaining after error correction (detection) has been performed. This setting should be distinguished from recent work on physical-level MSD \cite{ruiz_unfolded_2025,brown_advances_2023}, where distillation is performed directly on physical magic states before they are encoded into logical magic states.

From the perspective of mapped dynamical system, the measurement threshold is a dynamical transition of the distillation map characterized by a saddle-node bifurcation: below the critical measurement strength $\beta_{\mathrm{c}}$ no attractive distilling fixed point exists even for perfect inputs, whereas above $\beta_{\mathrm{c}}$ the map converges to biased fixed points. The critical behavior occurs when the largest eigenvalue of the Jacobian matrix at the fixed point reaches to 1 (see Supplementary Note~\ref{app:jacobian}). This mechanism is distinct from the canonical known input-fidelity threshold \cite{bravyiUniversalQuantumComputation2005, reichardtImprovedMagicStates2005, campbell_magic-state_2012, meierMagicstateDistillationFourqubit2012}, which only concerns the initial conditions for a fixed, ideal (projective-measurement) map rather than a change in the dyanmical map’s phase portrait.

In the same noise model and standard-form generator description, The specific critical threshold for measurement strength will depend on the code parameter, the threshold value will be decided by the competition between the ability of the MSD protocols to dump noise out and the noise introduced by imperfect measurements. For a $[[n, k, d]]$ transversal MSD protocol, the ability to dump noise out is characterized by the code distance $d$, and the noise introduced by imperfect measurement should scale with $n-k$. The noisier the protocol is, the higher the threshold is for the measurement strength. Therefore, we predict that the threshold value $\beta_*$  should be relate to $(n-k)/d$. For "good" distillation protocols with $d\propto O(n)$, the threshold for this family should converge to a fixed value when $n$ scales up. 

Given the impact of imperfect measurements, we might need to develop techniques to further mitigate imperfect measurement noise in practical regime. In the logical level, similar to the spirit of single-shot error correction \cite{kubica_single-shot_2022, delfosse_beyond_2022}, we could also measure redundant stabilizer elements beyond the stabilizer generators.  For hardware where physical qubits can be measured non-destructively, the imperfect measurement noise can be mitigated by repetitive physical measurement \cite{maHighfidelityGatesMidcircuit2023a, graham_mid-circuit_2023,bluvstein_architectural_2025}. We leave this direction as potential future effort.

\section*{Methods}

\subsection*{Simulating distillation under imperfect measurements}
\label{met: 1}
In the ideal measurement case, magic state distillation can be simulated and visualized by mapping to dynamical systems \cite{zhengMagicStateDistillation2024}. In particular for single-qubit magic states, the dynamical systems is a rational function defined on the Bloch sphere: $\mathcal{D}(x,y,z)$. Consider an $n$-to-$1$ protocol described by $\mathcal{Q}$, the dynamical system is given by
\begin{equation}
\mathcal{D}(x,y,z) =
    \begin{cases}
        x^o = T_{\mathrm{x}}(x,y,z)\\
        y^o = T_{\mathrm{y}}(x,y,z)\\
        z^o = T_{\mathrm{z}}(x,y,z).
    \end{cases}
    \label{eq: output_coord}
\end{equation}
$T_{\mathrm{x}}, T_{\mathrm{y}}, T_{\mathrm{z}}$ are all rational function. For instance, The explicit expression for $T_{\mathrm{x}}(x,y,z)$ is given by
\begin{equation}
        T_{\mathrm{x}} = \dfrac{\sum_{j=1}^{2^{\bar n}}(-1)^{\alpha_j}x^{\Tilde{w}^X_{ j}}y^{\Tilde{w}^Y_{j}}z^{\Tilde{w}^Z_{j}}}{\sum_{j=1}^{2^{ \bar n }}(-1)^{\alpha_j}x^{w^X_j}y^{w^Y_j}z^{w^Z_j}},
\end{equation}
where the sum is over all possible stabilizer element in $\mathcal{Q}$. $\alpha_j$ is the sign for the $j$-th element, and $w^P_j$ ($\tilde w^P_j$) is the Pauli weight of symbol $P\in\{X,Y,Z\}$ for the $j$-th element $s_j$ ($s_j * \bar{X}$).

To incorporate the effect of imperfect measurements in MSD, we instead need to adopt the mapping method to accommodate the impact of imperfect measurements. As we show in Eq. \eqref{eq:projector} and Eq. \eqref{eq:m_square}, the difference between the imperfect and ideal case is the coefficient. Therefore, we need to accommodate the coefficient change and reflect it in the explicit expression. In such a case, $T_{\mathrm{x}}$ is given by

\begin{equation}
        T_{\mathrm{x}} = \dfrac{\sum_{j=1}^{2^{\bar n}}(-1)^{\alpha_j} (h^2(\beta))^{\gamma_j} x^{\Tilde{w}^X_{ j}}y^{\Tilde{w}^Y_{j}}z^{\Tilde{w}^Z_{j}}}{\sum_{j=1}^{2^{ \bar n }}(-1)^{\alpha_j} (h^2(\beta))^{\gamma_j} x^{w^X_j}y^{w^Y_j}z^{w^Z_j}},
\end{equation}
where $\gamma_j$ is the number of generators required to generate the element $s_j$ and $h(\beta)$ is the coefficient that describes the measurement imperfection. $T_{\mathrm{y}}, T_{\mathrm{z}}$ in the imperfect case can be obtained in the similar way. 

\subsection*{Estimating asymptotic distillation cost}
\label{met: 2}
The distillation cost for magic states is defined to be the number of raw magic states required to prepare an $\epsilon$-good output state.
For an $n$-to-$k$ MSD protocols that suppress error with order-$d$ ($d>2$), i.e., the output error  $\epsilon_{\mathrm{o}}=k'\epsilon_{\mathrm{i}}^d$ with constant $k'$ and input error $\epsilon_{\mathrm{i}}$, the cost $C(\epsilon)$ is propotional to $(\log{\epsilon^{-1}})^\gamma$, where $\gamma = \log(n/k)/\log(d)$ is a protocol-specific parameter. To see that, after $l$-level iteration with raw input error $\epsilon_{\mathrm{r}}$, the output error is
\begin{equation}
    \epsilon_{\mathrm{o}} = k'^{\frac{d^l-1}{d-1}}\epsilon_{\mathrm{r}}^{d^l}.
\end{equation}
Let $\epsilon_{\mathrm{o}} = \epsilon$, we have 
\begin{equation}
    l = \dfrac{\log{\frac{\log{\epsilon k'^{1/(d-1)} }}{\log{\epsilon_{\mathrm{r}} k'^{1/(d-1)}}}}}{\log{d}}.
\end{equation}
The cost $C(\epsilon)= (n/k)^l$, so we have
\begin{equation}
C(\epsilon)
=
\left[
\frac{\log\left(\epsilon k'^{1/(d-1)}\right)}
{\log\left(\epsilon_{\mathrm{r}} k'^{1/(d-1)}\right)}
\right]^{
\frac{\log(n/k)}{\log d}
}\propto (\log{\epsilon^{-1}})^\gamma.
\end{equation}
Here both the numerator and denominator inside the bracket are negative in the relevant regime, so the ratio is positive and scales as $\log(\epsilon^{-1})$ in the asymptotic limit.
For example, the cost scaling for the $[[15, 1, 3]]$ protocol is approxiamtely $\log^{2.46}(1/\epsilon)$.

However, the scaling no longer holds for when error suppression is only linear. Consider another $n$-to-$k$ protocol with linear error suppression, i.e. $\epsilon_{\mathrm{o}}=k'\epsilon_{\mathrm{i}}$ with $k'<1$, the cost scaling is given by $(\epsilon_{\mathrm{r}}/\epsilon)^{\tau}$. $\epsilon_{\mathrm{r}}$ is the error rate of raw states and $\tau = \log(n/k)/\log(1/k')$. 

For completeness we provide the full derivation. We consider a nested MSD protocol where the output state will be feed as the input of the next level distillation. At the level-$l$, the total raw states cost is $(n/k)^l$, and the output error rate is $k'^l \epsilon_{\mathrm{r}}$. In the asymptotic limit, the distillation level $l^*$ should satisfy
\begin{equation}
    k'^{l^*} \epsilon_{\mathrm{r}} = \epsilon
\end{equation}
to fulfill the error requirement. The solution is $l^* = \log(\epsilon/\epsilon_{\mathrm{r}})/\log(k')$. Therefore, the cost $C_{\mathrm{linear}}(\epsilon)$ is given by
\begin{equation}
\begin{split}
      C_{\mathrm{linear}}(\epsilon) &= (n/k)^{l^*} = (\epsilon/\epsilon_{\mathrm{r}})^{1/\log_{n/k}k'} \\
      &=(\epsilon_{\mathrm{r}}/\epsilon)^\tau.
\end{split}
\end{equation}

There are two main caveats when the distillation efficiency is only linear. First, the distillation cost $C_{\mathrm{linear}}(\epsilon)$ is exponentially higher than $C(\epsilon)$. Therefore, more resources are required to distill the same output states in the asymptotic limit. Second, the cost function now explicitly depends on the quality of the raw states. In comparison, the cost function for $d>2$ doesn't explicitly depend on the raw error rate as long as it is above the distillation threshold \cite{bravyiUniversalQuantumComputation2005}. 
\\

\section*{Data Availability}
No experimental data were used in this work. The datasets generated and/or analyzed during the current study are not publicly available because no public repository has been established for this study, but are available from the corresponding author on reasonable request.

\section*{Code Availability}
The custom code generated and/or analyzed during the current study is not publicly available because no public repository has been established for this study, but is available from the corresponding author on reasonable request.

\section*{Acknowledgements}
We thank Ming Yuan for valuable comments on the motivation of the work. Yunzhe Zheng thanks Pei-Kai Tsai for insightful discussion on the proof of the theorem. We thank Bailey Gu for valuable comments on the manuscript. This work was supported by the National Natural Science Foundation of China (Grant No.~92365111), Shanghai Municipal Science and Technology (Grant No.~25LZ2600200), Beijing Natural Science Foundation (Grant No.~Z220002), and Quantum Science and Technology-National Science and Technology Major Project (Grant No.~2021ZD0302400). The funders had no role in study design, data collection and analysis, decision to publish, or preparation of the manuscript.

\section*{Author Contributions}
Y. Zheng and D.E. Liu conceived the project. Y. Zheng and Y. Zhao carried out the analytical calculations, and Y. Zheng performed the numerical simulations under the supervision of D.E. Liu. All authors discussed the results and contributed to writing the manuscript. All authors read and approved the manuscript.

\section*{Competing Interests}
The authors declare no competing financial or non-financial interests.

\clearpage
\onecolumngrid
\setcounter{section}{0}
\setcounter{figure}{0}
\setcounter{table}{0}
\setcounter{equation}{0}
\setcounter{definition}{0}
\setcounter{lemma}{0}
\setcounter{theorem}{0}
\setcounter{corollary}{0}
\setcounter{proposition}{0}
\setcounter{conjecture}{0}
\setcounter{example}{0}
\renewcommand{\thesection}{S\arabic{section}}
\renewcommand{\thefigure}{S\arabic{figure}}
\renewcommand{\thetable}{S\arabic{table}}
\renewcommand{\theequation}{S\arabic{equation}}
\renewcommand{\thedefinition}{S\arabic{definition}}
\renewcommand{\thelemma}{S\arabic{lemma}}
\renewcommand{\thetheorem}{S\arabic{theorem}}
\renewcommand{\thecorollary}{S\arabic{corollary}}
\renewcommand{\theproposition}{S\arabic{proposition}}
\renewcommand{\theconjecture}{S\arabic{conjecture}}
\renewcommand{\theexample}{S\arabic{example}}
\renewcommand{\theHsection}{supp.\arabic{section}}
\renewcommand{\theHfigure}{supp.\arabic{figure}}
\renewcommand{\theHtable}{supp.\arabic{table}}
\renewcommand{\theHequation}{supp.\arabic{equation}}

\begin{center}
{\large\bfseries Supplementary Information for\\[0.4em]
``Fragility of Magic State Distillation under Imperfect Measurements''\par}
\vspace{1em}
Yunzhe Zheng, Yuanchen Zhao, and Dong E. Liu
\end{center}

\tableofcontents

\section{Supplementary Note 1: Unnecessity for encoding in practical MSD}
\label{app:encoding}
It's unnecessary to encode the input states into a logical state in the framework of stabilizer reduction \cite{zhengMagicStateDistillation2024}. The conclusion holds for all stabilizer codes with transversal non-Clifford gates, but they only provide MSD protocols distilling into $\theta_{k}=  (\ket{0}+e^{i\pi/2^k}\ket{1})/\sqrt{2}$ (or their Clifford equivalence) due to the restriction on transversal logical gates \cite{andersonClassificationTransversalGates2014}. 

\begin{figure*}[t!]
    \centering
    \includegraphics[width=\textwidth]{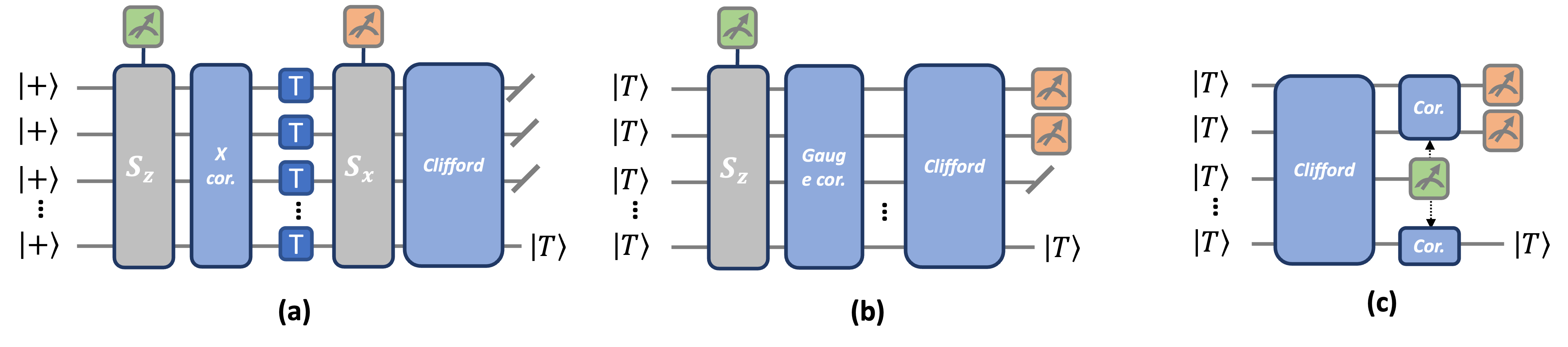}
    \caption{Equivalence between code with transversal $T$ gates and a stabilizer reduction protocol for magic state $\ket{T}$. (a) Distillation circuit with logical state preparation. We prepare the logical $\ket{+}_{\mathrm{L}}$ by preparing tensor product of $\ket{+}^{\otimes n}$ and measuring all $Z$-type stabilizer generators. We then apply $\bar{T} = T^{\otimes 15}$ and decode it with error detection.  (b) As $T^{\otimes n}$ commutes with the state preparation operator $\bar{P}_{\mathrm{Z}}$, we can bring it forward and act on the $\ket{+}^{\otimes n}$ directly. (c) We can then propagate the stabilizer measurement after the decoder circuit, which will be exactly measuring ancilla qubits. }
    \label{fig:equivalence_T_code}
\end{figure*}

To see this fact, we take the protocols for the $\ket{T}$ states and start with the canonical way for error corrected $T$ gates as Figure \ref{fig:equivalence_T_code}(a). We prepare the logical $\ket{+}_{\mathrm{L}}$ by initializing tensor product state $\ket{+}^{\otimes n}$ and measure every $Z$-type stabilizer generators. Gauge correction might be required depending on the measurement outcome. This preparation operation can be characterized by the projector $\bar{P}_{\mathrm{Z}} = \prod_{ g\in\mathcal{G}_{\mathrm{Z}}}\frac{I+g}{2}$ where $\mathcal{G}_{\mathrm{Z}}$ is the set of all $Z$-type stabilizer generators. As show later in Lemma \ref{lemma: commutation}, it commutes with the transversal $T$ operations. Therefore, we can move the transversal $T$ gates to the very beginning and merge the stabilizer measurements for $X$-type and $Z$-type stabilizer as Figure \ref{fig:equivalence_T_code}(b). Lastly, we should notice that the decoder operation convert the logical state in the codespace of $\mathcal{Q}$ back to physical state, so it will also propagate stabilizer generators back to the raw physical $Z$ operators on every ancilla qubits. We can then postpone the measurement after decoding. The gauge correction can also be propagated after the decoder operation as it's a Clifford operation, but we only need to track the effect of propagated gauge correction on the decoded data qubits and applying it back to our physical state, as shown in Figure \ref{fig:equivalence_T_code}(c).

We emphasize that these gauge corrections are not needed in the stabilizer-reduction (SR) description used in the main text. The discussion above only shows the equivalence between the usual deterministic logical-state-preparation circuit and the SR circuit. In the SR protocol, the relevant map is the post-selected branch in which the measured stabilizer generators return the accepted, all-$+1$ syndrome. Therefore, no gauge correction is applied in the SR analysis. The subspace-transfer operators that appear later in the proofs should instead be understood as algebraic Pauli operators that relate a syndrome-subspace projector to the codespace projector, e.g., $P_{\mathbf{x}}=C_{\mathbf{x}}\bar P C_{\mathbf{x}}$, rather than as feed-forward gauge corrections performed by the protocol. Consequently, including or omitting gauge corrections in the circuit representation does not affect the noisy-projector analysis or the standard-form mitigation strategy.

\section{Supplementary Note 2: Description of $[[15, 1, 3]]$ and $[[14, 2, 2]]$ codes}
\label{app: code_description}
Both the $[[15, 1, 3]]$ \cite{bravyiUniversalQuantumComputation2005} and $[[14, 2, 2]]$ \cite{bravyiMagicstateDistillationLow2012} codes have logical transversal $T$ gates and therefore can be used as MSD protocol to distill the $T$ states. We give their stabilizer description here as they are practical small examples to work with. 

\begin{figure}[t!]
    \centering
    \includegraphics[width=\linewidth]{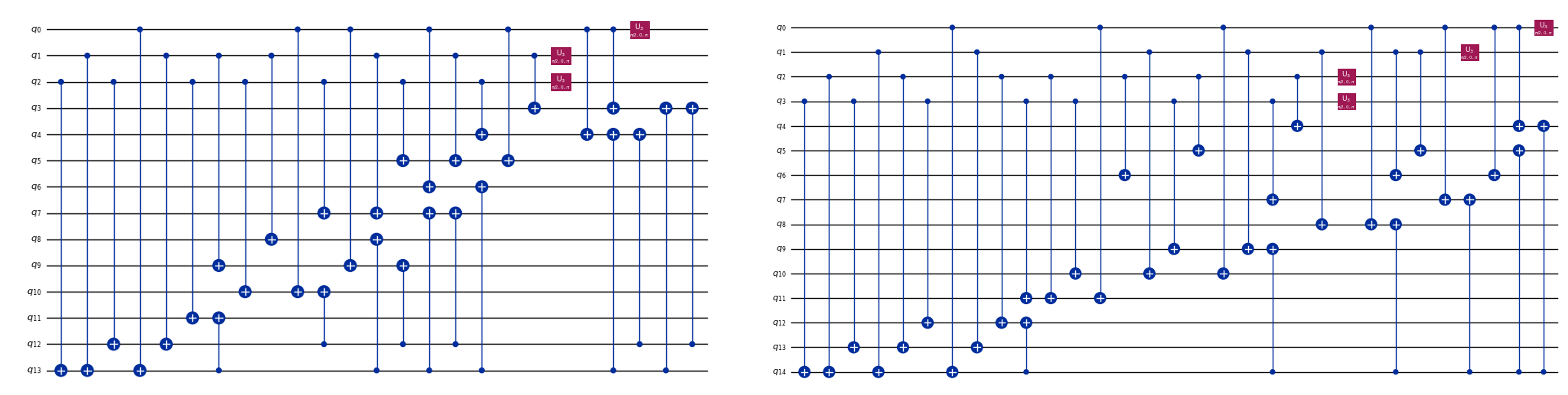}
    \caption{\textbf{Decoding circuits for the two example MSD protocols.} The left panel shows the decoding circuit for the $[[15, 1, 3]]$ protocol and the right panel shows the decoding circuit for the $[[14, 2, 2]]$ protocol. The output qubits are placed at the bottom, and all other qubits are measured in the $Z$ basis.}
    \label{fig:decoding_circuits_app}
\end{figure}

For concreteness, Figure~\ref{fig:decoding_circuits_app} shows decoding circuits for the two protocols in the stabilizer-reduction implementation used in the main text.

The $[[15, 1, 3]]$ code has 10 $Z$-type generators and 4 $X$-type generators. For the canonical generator choice, the parity check matrix is given by
\begin{equation}
    H_{\mathrm{X}} = \begin{bmatrix}
    1 & 0 & 1 & 0 & 1 & 0 & 1 & 0 & 1 & 0 & 1 & 0 & 1 & 0 & 1\\ 
    0 & 1 & 1 & 0 & 0 & 1 & 1 & 0 & 0 & 1 & 1 & 0 & 0 & 1 & 1\\  
    0 & 0 & 0 & 1 & 1 & 1 & 1 & 0 & 0 & 0 & 0 & 1 & 1 & 1 & 1\\ 
    0 & 0 & 0 & 0 & 0 & 0 & 0 & 1 & 1 & 1 & 1 & 1 & 1 & 1 & 1
    \end{bmatrix} \ \ 
    H_{\mathrm{Z}} = \begin{bmatrix}
    1 & 0 & 1 & 0 & 1 & 0 & 1 & 0 & 1 & 0 & 1 & 0 & 1 & 0 & 1\\ 
    0 & 1 & 1 & 0 & 0 & 1 & 1 & 0 & 0 & 1 & 1 & 0 & 0 & 1 & 1\\  
    0 & 0 & 0 & 1 & 1 & 1 & 1 & 0 & 0 & 0 & 0 & 1 & 1 & 1 & 1\\ 
    0 & 0 & 0 & 0 & 0 & 0 & 0 & 1 & 1 & 1 & 1 & 1 & 1 & 1 & 1\\
    0 & 0 & 1 & 0 & 0 & 0 & 1 & 0 & 0 & 0 & 1 & 0 & 0 & 0 & 1\\
    0 & 0 & 0 & 0 & 1 & 0 & 1 & 0 & 0 & 0 & 0 & 0 & 1 & 0 & 1\\
    0 & 0 & 0 & 0 & 0 & 1 & 1 & 0 & 0 & 0 & 0 & 0 & 0 & 1 & 1\\
    0 & 0 & 0 & 0 & 0 & 0 & 0 & 0 & 0 & 1 & 1 & 0 & 0 & 1 & 1\\
    0 & 0 & 0 & 0 & 0 & 0 & 0 & 0 & 0 & 0 & 0 & 1 & 1 & 1 & 1\\
    0 & 0 & 0 & 0 & 0 & 0 & 0 & 0 & 1 & 0 & 1 & 0 & 1 & 0 & 1\\
    \end{bmatrix},
\end{equation}
Notably, $H_{\mathrm{Z}} = [H_{\mathrm{X}}, H_{\mathrm{Z}}']^T$. The logical operator is respectively $X_{\mathrm{L}} = X^{\otimes 15}$, $Z_{\mathrm{L}} = Z^{\otimes 15}$. For the parity-check matrices in standard form, we have
\begin{equation}
\label{eq:15_standard_form}
    H_{\mathrm{X}} = \begin{bmatrix}
    1 & 0 & 0 & 0 & 1 & 0 & 1 & 1 & 1 & 0 & 1 & 1 & 0 & 0 & 1\\ 
    0 & 1 & 0 & 0 & 0 & 1 & 1 & 0 & 1 & 1 & 1 & 0 & 0 & 1 & 1\\  
    0 & 0 & 1 & 0 & 1 & 1 & 1 & 0 & 0 & 0 & 0 & 1 & 1 & 1 & 1\\ 
    0 & 0 & 0 & 1 & 0 & 0 & 0 & 1 & 0 & 1 & 1 & 1 & 1 & 1 & 1
    \end{bmatrix} \ \ 
    H_{\mathrm{Z}} = \begin{bmatrix}
    0 & 1 & 0 & 1 & 1 & 0 & 0 & 0 & 0 & 0 & 0 & 0 & 0 & 0 & 1\\ 
    1 & 0 & 0 & 1 & 0 & 1 & 0 & 0 & 0 & 0 & 0 & 0 & 0 & 0 & 1\\  
    1 & 1 & 1 & 0 & 0 & 0 & 1 & 0 & 0 & 0 & 0 & 0 & 0 & 0 & 0\\ 
    0 & 1 & 1 & 0 & 0 & 0 & 0 & 1 & 0 & 0 & 0 & 0 & 0 & 0 & 1\\
    0 & 0 & 1 & 1 & 0 & 0 & 0 & 0 & 1 & 0 & 0 & 0 & 0 & 0 & 1\\
    1 & 0 & 1 & 0 & 0 & 0 & 0 & 0 & 0 & 1 & 0 & 0 & 0 & 0 & 1\\
    1 & 1 & 0 & 1 & 0 & 0 & 0 & 0 & 0 & 0 & 1 & 0 & 0 & 0 & 0\\
    1 & 0 & 1 & 1 & 0 & 0 & 0 & 0 & 0 & 0 & 0 & 1 & 0 & 0 & 0\\
    1 & 1 & 0 & 0 & 0 & 0 & 0 & 0 & 0 & 0 & 0 & 0 & 1 & 0 & 1\\
    0 & 1 & 1 & 1 & 0 & 0 & 0 & 0 & 0 & 0 & 0 & 0 & 0 & 1 & 0\\
    \end{bmatrix},
\end{equation}
and the logical operator is given by $X_{\mathrm{L}} = IIIIXXIXXXIIXIX$ and $Z_{\mathrm{L}} = ZZZZIIIIIIIIIIZ$.

The $[[14, 2, 2]]$ code has 9 $Z$-type generators and 3 $X$-type generators. For the canonical generator choice, it's parity check matrix is given by
\begin{equation}
    H_{\mathrm{X}} = \begin{bmatrix} 1 & 0 & 1 & 0 & 1 & 0 & 1  &1 & 0 & 1 & 0 & 1 & 0 & 1 \\ 0 & 1& 1&0&0&1&1 & 0 & 1& 1&0&0&1&1 \\ 0&0&0&1&1&1&1 &0&0&0&1&1&1&1\end{bmatrix} \ \ 
    H_{\mathrm{Z}} = \begin{bmatrix}  0 & 1 & 1 & 1 & 1 & 0 & 0 & 0 & 0 & 0 & 0 & 0 & 0 & 0\\  1 & 0 & 1 & 1 & 0 & 1 & 0 & 0 & 0 & 0 & 0 & 0 & 0 & 0\\  1 & 1 & 0 & 1 & 0 & 0 & 1 & 0 & 0 & 0 & 0 & 0 & 0 & 0\\  1 & 1 & 0 & 0 & 0 & 0 & 0 & 1 & 1 & 0 & 0 & 0 & 0 & 0\\  1 & 0 & 1 & 0 & 0 & 0 & 0 & 1 & 0 & 1 & 0 & 0 & 0 & 0\\  1 & 0 & 0 & 1 & 0 & 0 & 0 & 1 & 0 & 0 & 1 & 0 & 0 & 0\\  0 & 1 & 1 & 0 & 0 & 0 & 0 & 0 & 0 & 0 & 1 & 1 & 0 & 0\\  0 & 0 & 1 & 1 & 0 & 0 & 0 & 1 & 0 & 0 & 0 & 0 & 1 & 0\\  0 & 1 & 0 & 1 & 0 & 0 & 0 & 1 & 0 & 0 & 0 & 0 & 0 & 1 \end{bmatrix},
\end{equation}
The logical operators are respectively $X_{\mathrm{L},1} = XXXXXXXIIIIIII, Z_{\mathrm{L},1} = ZZZZZZZIIIIIII$, and $X_{\mathrm{L},2} = IIIIIIIXXXXXXX, Z_{\mathrm{L},2} = IIIIIIIZZZZZZZ$. For the parity-check matrices in standard form,
\begin{equation}
    H_{\mathrm{X}} = \begin{bmatrix}
    1& 0& 0& 1& 1& 1& 1& 0& 0& 1& 1& 0& 0& 1\\
    0& 1& 0& 1& 0& 1& 0& 1& 1& 1& 0& 0& 1& 1\\
    0& 0& 1& 0& 1& 1& 0& 1& 0& 0& 1& 1& 1& 1
    \end{bmatrix} \ \
    H_{\mathrm{Z}} = \begin{bmatrix}
    0& 1& 0& 1& 0& 0& 0& 0& 0& 0& 0& 0& 1& 1\\
    0& 0& 1& 0& 1& 0& 0& 0& 0& 0& 0& 0& 1& 1\\
    1& 1& 1& 0& 0& 1& 0& 0& 0& 0& 0& 0& 0& 0\\
    0& 1& 1& 0& 0& 0& 1& 0& 0& 0& 0& 0& 0& 1\\
    1& 1& 1& 0& 0& 0& 0& 1& 0& 0& 0& 0& 1& 1\\
    1& 0& 1& 0& 0& 0& 0& 0& 1& 0& 0& 0& 0& 1\\
    1& 0& 1& 0& 0& 0& 0& 0& 0& 1& 0& 0& 1& 0\\
    1& 1& 0& 0& 0& 0& 0& 0& 0& 0& 1& 0& 1& 0\\
    1& 1& 0& 0& 0& 0& 0& 0& 0& 0& 0& 1& 0& 1
    \end{bmatrix},
\end{equation}  
and the logical operators are respectively $X_{\mathrm{L},1}=IIIIIIXIXXXXXX, Z_{\mathrm{L},1} = IZZIIIIIIIIIZI$ and $X_{\mathrm{L},2} = IIIXXIXXXIIXIX, Z_{\mathrm{L},2} = ZIIIIIIIIIIIZZ$. 

\section{Supplementary Note 3: Gaussian measurement error model and generality}
\label{app:gaussian}

In the main text, we considered the discretized noisy measurement operators under Gaussian noise. In practice, the measurement outcome is continuous and the noisy measurement under Gaussian noise for observable $g$ with outcome $s$ is given by:
  \begin{equation}
      E(s) = (k/2\pi)^{1/2} \exp\left( -k (s-g)^2 /2\right), \quad \int^{+\infty}_{-\infty} ds E(s) = I,
  \end{equation}
and $E(s) = M^\dagger(s) M(s)$ where $M(s)$ is the weak measurement operator. The outcome provided by the detector is a real number $s$, but since the eigenvalue of $g$ is $\pm 1$ and we expect a binary outcome, in practice we often identify the $s>0$ outcome as $\tilde{s}=+1$ and $s<0$ outcomes as $\tilde{s} = -1$. This leads to the measurement probability,
\begin{equation}
    P(\tilde{s}=\pm 1) = P(\pm s > 0) = \frac{1}{2} \left(1 \pm g \Erf \left(\sqrt{k/2}\right)\right),
\end{equation}
where $\Erf$ is the Gaussian error function. This probability is equivalent to that yielded by the binary model used in the simulation
\begin{equation}
    E(\tilde{s}) \propto e^{- \tilde{s} \beta g},
\end{equation}
where we defined the measurement strength $\beta = \arctanh (\Erf (\sqrt{k/2}))$.

In the generic case, the format of $E(s)$ doesn't need to be Gaussian. However, whatever it turns to be, the discretized measurement operator will still be a similar form as the Gaussian case, except for a change of the coefficients. Particularly, the noisy measurement operator for a qubit Pauli observable $g$ is given by

\begin{equation}
    \tilde{M}_+ = f_1P_+ + f_2P_-, \quad \tilde{M}_- = h_1P_- + h_2P_+,
\end{equation}
where $P_\pm = \frac{I\pm g}{2}$ is the projector for $\pm 1$ eigenspace of $g$. The the normalization condition $\tilde{M}^\dagger_+ \tilde{M}_+ + \tilde{M}^{\dagger}_- \tilde{M}_- = I$ demands 
\begin{equation}
    f_1^2 + h_2^2 = 1, \quad f_2 ^2 + h_1 ^2 = 1,
\end{equation}
and the semi-positive property for measurement operators requires $f_{1,2}$ and $h_{1,2}$ to be non-negative. If we are under such a noise model, we can still use the conclusion and analysis for Gaussian noise by a parameter mapping. The mapping is given by 
\begin{equation}
    f_2/f_1 = \tanh{\beta/2},
\end{equation}
and we can relate the new defined noise parameter with the measurement strength for gaussian noise. For example, we may consider the following noisy measurements that differ from the Gaussian noise:
\begin{equation}
    \tilde{M}_+ = \sqrt{\frac{1+\eta}{2}} P_+ + \sqrt{\frac{1-\eta}{2}} P_-, \quad \tilde{M}_- = \sqrt{\frac{1-\eta}{2}}P_+ + \sqrt{\frac{1+\eta}{2}} P_-,
    \label{eq: meas_model}
\end{equation}
where $\eta \in [0, 1]$ now is the new "measurement strength". We then have the equation
\begin{equation}
    \sqrt{\frac{1-\eta}{1+\eta}} = \tanh \beta/2.
\end{equation}

Now all the analysis showed with $\beta$ can be showed using $\eta$ by replacing the parameter using the above relation. For example, the first-order biased noise for the $[[15, 1, 3]]$ protocol now scales as $e^{-\tanh^{-1}(\sqrt{\frac{1-\eta}{1+\eta}})}$ rather than $e^{-2\beta}$.

\section{Supplementary Note 4: Used Lemmas}
\label{app:lemmas}
\begin{lemma}
    If a $n$-qubit stabilizer code $Q$ allows transversal implementation of logical $K$ gates, then $K^{\otimes n}$ commutes with the codespace projector $\bar{P}$.
    \label{lemma: commutation}
\end{lemma}
\begin{proof}
For any $n$-qubit quantum states $\ket{\psi}$ and stabilizer code $\mathcal{Q}$ with codespace projector $\bar{P}$, we can decompose the state into components in/out of the codespace defined by $\mathcal{Q}$ as $\ket{\psi} = \ket{\psi}_{\mathrm{i}}+\ket{\psi}_{\mathrm{o}}$  such that $\bar{P}\ket{\psi} = \ket{\psi}_{\mathrm{i}}, (I-\bar{P})\ket{\psi} = \ket{\psi}_{\mathrm{o}}$. As logical operations should preserve the codespace as well as the non-codespace and $K^{\otimes n}$ is a logical operation, $\bar{P}K^{\otimes n}\ket{\psi}_0 = 0$. Therefore,
\begin{equation}
    \bar{P}K^{\otimes n} \ket{\psi} = \bar{P} (K^{\otimes n}\ket{\psi}_{\mathrm{i}}+K^{\otimes n}\ket{\psi}_{\mathrm{o}}) =K^{\otimes n} \ket{\psi}_{\mathrm{i}}.
\end{equation}
Besides, we can easily have
\begin{equation}
    K^{\otimes n}\bar{P} \ket{\psi} = K^{\otimes n} \ket{\psi}_{\mathrm{i}}.
\end{equation}
As $K^{\otimes n}\bar{P} \ket{\psi} = \bar{P}K^{\otimes n} \ket{\psi}$ for any quantum states, $\bar P$ must commute with $K^{\otimes n}$.
\end{proof}

Although we don't have any restriction for the transversal $K$ gates, the only possible transversal gates are just $R_{\mathrm{Z}}(\theta)$ gates (or their Pauli equivalence), with $\theta = \pi/2^k$ for $k$ being a non-negative integer  \cite{zengTransversalityUniversalityAdditive2007,andersonClassificationTransversalGates2014}. 
\begin{lemma}
For a CSS stabilizer code $\mathcal{Q}$ with generator $\{g_i\}$, we can always find a pure X or Z Pauli subspace-transfer operator $C_i$, i.e. composed of only one type of Pauli operators ($X,Y$ or $Z$) and $I$, to map its codespace to another subspace that is stabilized by $\{(-1)^{\alpha_i}g_i\}$ where $\alpha_i \in \{0, 1\}$ and $|\bm{\alpha}|=1$ with $\bm{\alpha}=\{\alpha_1, \alpha_2...\}$, while preserving the logical state.
\label{lemma: corr}
\end{lemma}
\begin{proof}
Without loss of generality we consider the subspace $\mathcal{\Tilde{Q}}$ where only the syndrome of $X$-type stabilizer $g_k$ is flipped. The subspace projector associated with $\mathcal{\Tilde{Q}}$ is now:
\begin{equation}
    P_{\mathcal{\tilde{Q}}} =  \frac{I-g_k}{2} \prod^{\bar{n}}_{j=1,j\neq k} \frac{I+g_j}{2}.
\end{equation}
As the subspace-transfer operator $C_k$ should map the codespace to the subspace, we have $C_k\bar{P}C_k  = P_{\mathcal{\tilde{Q}}} =  \frac{I-g_k}{2} \prod^{\bar{n}}_{j=1,j\neq k} \frac{I+g_j}{2}$. Therefore, $C_k$ can be understood as a physical error that only triggers the $g_k$ syndrome and we can restrict it to be $Z$-type errors. Furthermore, $C_k$ will preserve logical state and therefore should commute with all logical operators. For CSS codes, we can choose logical $X(Z)$ to be pure $X(Z)$ operators. If $C_k$ commutes with all logical $X$ operators, then we are all set and the $C_k$ is the given subspace-transfer operator. If not, we can multiply $C_k$ with the logical $Z$ operators associated with the non-commuting local $X$ operators as the $\tilde{C}_k$. The $\tilde{C}_k$ can then be the subspace-transfer operator we want.
\end{proof}

\begin{lemma}
    For a stabilizer code whose generators are written in standard form, the subspace-transfer operators $C_i$ in Lemma \ref{lemma: corr} can be chosen to have weight one, to have pairwise disjoint support, and to commute with all logical operators.
    \label{lem: standard_corr}
\end{lemma}
\begin{proof}
Write the stabilizer generators in standard form
\begin{equation}
    [g_1, g_2, ... ,g_{\bar{n}}]^T =  [H_{\mathrm{X}} | H_{\mathrm{Z}}] = \begin{bmatrix}[ccc|ccc]
        I & A_1 & A_2 & B & 0 & C \\
        0 & 0 & 0 & D & I & E
    \end{bmatrix}.
\end{equation}
Here $r=\operatorname{rank}(H_{\mathrm{X}})$. For the first $r$ generators, choose $C_i=Z_i$. For the remaining generators, indexed as $g_{r+a}$ with $1\leq a\leq \bar n-r$, choose $C_{r+a}=X_{r+a}$. By the identity blocks in this standard-form representation, each chosen $C_i$ anticommutes only with the corresponding generator $g_i$ and commutes with all other generators. Hence $C_i$ maps the codespace to the syndrome subspace in which only the $i$-th syndrome is flipped. The operators $C_i$ are manifestly weight-one and have pairwise disjoint support.

The logical operators associated with this standard form may be chosen as \cite{nielsenQuantumComputationQuantum2010}
\begin{equation}
    L_{\mathrm{X}} = \begin{bmatrix}[ccc|ccc]
        0 & E^T & I & C^T & 0 & 0 
    \end{bmatrix},     L_{\mathrm{Z}} = \begin{bmatrix}[ccc|ccc]
        0 & 0 & 0 & A_2^T & 0 & I 
    \end{bmatrix}.
\end{equation}
The supports selected above lie in columns where these representatives have no anticommuting support. Therefore every $C_i$ commutes with both $L_{\mathrm{X}}$ and $L_{\mathrm{Z}}$, and hence with the logical Pauli group.
\end{proof}

For example, in the $[[15,1,3]]$ standard-form matrix in Equation~\eqref{eq:15_standard_form}, the four $X$-type generators have the identity block on the first four qubits, so one can choose the corresponding subspace-transfer operators as $Z_1,Z_2,Z_3,Z_4$. The ten $Z$-type generators have the identity block on qubits $5$ through $14$, so the corresponding subspace-transfer operators can be chosen as $X_5,\ldots,X_{14}$; these single-qubit operators have disjoint support and commute with the standard-form logical operators $X_{\mathrm{L}}=IIIIXXIXXXIIXIX$ and $Z_{\mathrm{L}}=ZZZZIIIIIIIIIIZ$.

\begin{lemma}
    The subspace-transfer operators in Lemma 2 can always be chosen such that they don't support logical operators.
    \label{lemma:support}
\end{lemma}
\begin{proof}
If the generators of the code are in standard form, then all subspace-transfer operators apparently don't support logical operators. Now let's consider an arbitrary CSS code with $X$-type and $Z$-type generators. Since the parity check matrix for a CSS code can always be converted into standard form by row adding and qubit permutation, we consider the effect of row adding on the subspace-transfer operator. Without loss of generality we consider $X$-type generator: As long as we just add $X$-type row with $X$-type row, the subspace-transfer operator will always support on the first $r$ qubits. However, no logical operators will be supported on the first $r$ qubits. Besides, qubit permutation doesn't modify the weight of subspace-transfer operators and logical operators.
\end{proof}

\section{Supplementary Note 5: Proof of Theorems}
\label{app:proofs}

\begin{theorem}
    For MSD protocols based on $[[n, k, d]]$ ($d\geq 2$) CSS codes with transversal non-Clifford gates, the first-order effect of weak measurement for $X$-type generators won't affect the asymptotically distilled output states up to the first order of $e^{-2\beta}$. The effect of weak measurement for $Z$-type generators will be at most $X$ noise on the asymptotically distilled output states up to the first order of $e^{-2\beta}$. Therefore, the first-order noise ($e^{-2\beta}$), if present, must be biased.
    \label{th: X_gen}
\end{theorem}
\begin{proof}
Without loss of generality, we may assume the Gaussian noise to clearly denote the noise order. (Note the proof doesn't rely on this property) The noisy measurement operator for stabilizer $g$ to measure +1 can be written as 
\begin{equation}
\begin{split}
        M_+(g,\beta) &= K(I+\tanh{\frac{\beta}{2}}g) = K(\frac{2}{1+e^{-\beta}}P_+ + \frac{2}{1+e^{\beta}}P_-),
\end{split}
\end{equation}
where $P_\pm = (I\pm g)/2$. We define an $\bar{n}$-bit binary string operator on $\mathbb{F}_2$: $\mathbf{x}=(x_1,x_2,\dots,x_{\bar{n}})$. For stabilizer code $\mathcal{Q}$ with $\bar{n}$ stabilizer generators, all the subspace projectors associated with stabilizer code $\mathcal{Q}$ are given by
\begin{equation}
    \bar{P}_{\mathbf{x}} = \frac{1}{2^{\bar n}}\prod^{\bar{n}}_{i=1}(I+(-1)^{x_i}g_i).
\end{equation}
It's easy to see we recover codespace projector 
$\bar{P}$ when $\mathbf{x}=\bf{0}$, i.e. $\bar{P}=\bar{P}_{\bf 0}$. Therefore, we can rewrite the codespace measurement operator as a sum of subspace projector:
\begin{equation}
    \tilde{M}_{\mathcal{Q}} = \prod^{\bar n}_{i=1} M_+(g_i, \beta) = K^{\bar n}\sum_{\mathbf{x}\in \mathbb{F}^2}\gamma_{\mathbf{x}}\bar{P}_{\mathbf{x}},
\end{equation}
where 
\begin{equation}
    \gamma_{\mathbf{x}} = (\frac{2e^{\beta}}{1+e^{\beta}})^{\bar{n}-|\mathbf{x}|}(\frac{2}{1+e^{\beta}})^{|\mathbf{x}|} = (\frac{2}{1+e^\beta})^{\bar n} e^{-\beta |\mathbf{x}|}.
\end{equation}
The post-measurement state of $\rho_{\mathrm{in}}$ is then given by
\begin{equation}
\begin{split}
        \rho_{\mathrm{p}} &\propto \tilde{M}_{\mathcal{Q}} \rho_{\mathrm{in}} \tilde{M}_{\mathcal{Q}} \\
        &\propto \sum_{\mathbf{x}, \mathbf{x'}}\gamma_{\bf{x}}\gamma_{\bf{x'}}\bar{P}_{\bf{x}} \rho_{\mathrm{in}} \bar{P}_{\bf{x'}}\\
        &\propto\sum_{\mathbf{x},\bf{x'}}e^{-(|\bf{x}|+|\bf{x'}|)\beta}\bar{P}_{\bf{x}} \rho_{\mathrm{in}} \bar{P}_{\bf{x'}}.
\end{split}
\end{equation}
Now we consider the final output state $\rho_{\mathrm{o}}$ decoded from the post-measurement state $\rho_{\mathrm{p}}$. There are two procedure in the decoding process for MSD: First, we should apply the decoding circuit \cite{zhengMagicStateDistillation2024} (the inverse of encoding circuit found using Gottesman's algorithm \cite{gottesmanStabilizerCodesQuantum1997}). Second, we should trace out all ancilla qubits and only keep these storing logical information. When the measurement is perfect, the logical qubits are fully unentangled with all other qubits and tracing out won't impact the output state. However, they are indeed entangled in the imperfect measurement case, and the tracing out operation will impact the qubits storing logical information. For simplicity we consider $k=1$, then 

\begin{equation}
\begin{split}
        \rho_{\mathrm{o}} &= \Tr_{\mathrm{A}}[D_\mathcal{Q} \rho_{\mathrm{p}} D^\dagger_\mathcal{Q}]\\
        &\propto\sum_{\mathbf{x},\bf{x'}}e^{-(|\bf{x}|+|\bf{x'}|)\beta}\Tr_{\mathrm{A}}[D_\mathcal{Q}\bar{P}_{\bf{x}} \rho_{\mathrm{in}} \bar{P}_{\bf{x'}}D^\dagger_\mathcal{Q}]\\
        &\propto\sum_{\bf{x}}e^{-2|\bf{x}|\beta}\Tr_{\mathrm{A}}[D_\mathcal{Q}\bar{P}_{\bf{x}} \rho_{\mathrm{in}} \bar{P}_{\bf{x}}D^\dagger_\mathcal{Q}]
\end{split}
\label{eq: general_MSD_weak_form}
\end{equation}
where $\Tr_{\mathrm{A}}$ means tracing out all ancillary qubits and from the second row to third row we used the fact that non-diagonal terms don't contribute when the ancillary qubits will be traced out after decoding the logical states to physical states. Let's restrict $\rho_{\mathrm{in}}=\rho^{\otimes n}_{\mathrm{i}}$ to be a product state. For leading-order consideration, we have
\begin{equation}
    \rho_{\mathrm{o}}(\rho_{\mathrm{i}}) \propto \Tr_{\mathrm{A}}[D_\mathcal{Q}\bar{P}_{\bf{0}} \rho^{\otimes n}_{\mathrm{i}} \bar{P}_{\bf{0}}D^\dagger_\mathcal{Q}] + \sum_{|\mathbf{x}|=1}e^{-2\beta}\Tr_{\mathrm{A}}[D_\mathcal{Q}\bar{P}_{\bf{x}} \rho^{\otimes n}_{\mathrm{i}} \bar{P}_{\bf{x}}D^\dagger_\mathcal{Q}].
    \label{eq: post_meas_state}
\end{equation}
Let's consider when $\rho_{\mathrm{in}}$ is the ideal input state. Say non-Clifford gate $R_{\mathrm{Z}}(\theta)$ is transversal for code $\mathcal{Q}$. The ideal input state can be written as
\begin{equation}
    \ket{\theta}^{\otimes n} = R_{\mathrm{Z}}(\theta)^{\otimes n}\ket{+}^{\otimes n}.
\end{equation}
The effect of codespace projector $\bar{P}_{\bf 0}$ on $\ket{\theta}^{\otimes n}$ is:
\begin{equation}
    \bar{P}_{\bf 0}\ket{\theta}^{\otimes n} =  \bar{P}_{\bf 0} R_{\mathrm{Z}}(\theta)^{\otimes n} \ket{+}^{\otimes n} = R_{\mathrm{Z}}(\theta)^{\otimes n}\bar{P}_{\bf 0}\ket{+}^{\otimes n} \propto \bar{R}_{\mathrm{Z}}(\theta)\ket{\overline +} = \ket{\overline \theta}.
\end{equation}
Therefore, the decoded state will be exactly $\ket{\theta}$, and the first term on in Equation \eqref{eq: post_meas_state} is proportional to $\ket{\theta}\bra{\theta}$. 

As the next step, we show the effect of $\bar{P}_{\bf{x}}$ on $\ket{\theta}^{\otimes n}$ for $|\mathbf{x}|=1$. Since the code is CSS code, the stabilizer generators should be either $X$-type or $Z$-type. If the subspace projector $\bar{P}_{g_{\mathrm{x}}}$ has syndrome -1 for an $X$-type generator $g_{\mathrm{x}}$ and +1 for all other stabilizers, then based on Lemma \ref{lemma: corr}, we can always find a subspace-transfer operator $\Tilde{Z}$ composed with only Pauli $Z$ that maps the codespace to this syndrome subspace without affecting the logical state, i.e., $\bar{P}_{g_{\mathrm{x}}} = \Tilde{Z}\bar{P}\Tilde{Z}$. Notably, both $\bar{P}_{g_{\mathrm{x}}}$ and $\Tilde{Z}$ depend on the choice of $g_{\mathrm{x}}$. However, $\bar{P}_{g_{\mathrm{x}}}\ket{\theta}^{\otimes n} = 0$. To see this,
\begin{equation}
\begin{split}
\bar{P}_{g_{\mathrm{x}}}\ket{\theta}^{\otimes n}   &= \Tilde{Z} \bar{P} \Tilde{Z} R_{\mathrm{Z}}(\theta)^{\otimes n}\ket{+}^{\otimes n}\\
&= \Tilde{Z} \bar{P}  R_{\mathrm{Z}}(\theta)^{\otimes n}\Tilde{Z}\ket{+}^{\otimes n}\\
&= \Tilde{Z}   R_{\mathrm{Z}}(\theta)^{\otimes n}\bar{P}\Tilde{Z}\ket{+}^{\otimes n}=0,
\label{eq: Z_type}
\end{split}
\end{equation}
where we used Lemma \ref{lemma: commutation} from the second row to the last row. The $\Tilde{Z}\ket{+}^{\otimes n}$ is a tensor product state of $\ket{+}$ and $\ket{-}$, while $\bar{P}$ contains projectors for every $X$-type stabilizer. As $\tilde{Z}$ should not be a logical operator, the state $\Tilde{Z}\ket{+}^{\otimes n}$ will be anti-stabilized by at least a $X$-type stabilizer, and $\bar{P}\Tilde{Z}\ket{+}^{\otimes n} = 0$. Therefore, $\bar{P}_{\bf{x}} (\ket{\theta} \bra{\theta})^{\otimes n} \bar{P}_{\bf{x}} = 0$ for $|\mathbf{x}|=1$ if the flipped stabilizer is $X$-type. 

Now we consider the case when $\bar{P}_{\bf{x}}$ has syndrome -1 for an arbitrary $Z$-type stabilizer $g_{\mathrm{z}}$. We denote the corresponding subspace projector by $\bar{P}_{g_{\mathrm{z}}}$, and $\Tilde{X}$ is the corresponding subspace-transfer operator between $\bar{P}_{g_{\mathrm{z}}}$ and $\bar{P}$ without affecting the logical state such that $\bar{P}_{g_{\mathrm{z}}}=\Tilde{X}\bar{P}\Tilde{X}$. We define the support of $\Tilde{X}$ as $\{m_i\}$. To calculate $\bar{P}_{g_{\mathrm{z}}}\ket{\theta}^{\otimes n}$, we define $Z_{m_i}$ as the Pauli operator $Z$ and $R_{\mathrm{Z},m_i}$ as the Pauli $Z$ rotation gate on qubit $m_i$:
\begin{equation}
\begin{split}
\bar{P}_{g_{\mathrm{z}}}\ket{\theta}^{\otimes n}   &= \Tilde{X} \bar{P} \Tilde{X} R_{\mathrm{Z}}(\theta)^{\otimes n}\ket{+}^{\otimes n}\\
&=\Tilde{X} \bar{P} (\bigotimes_{\{m_i\}} R_{\mathrm{Z}, m_i}(-2\theta)) R_{\mathrm{Z}}(\theta)^{\otimes n}\ket{+}^{\otimes n}\\
&=\Tilde{X} \bar{P} \bigotimes_{\{m_i\}} (\cos\theta I + i \sin\theta Z_{m_i}) R_{\mathrm{Z}}(\theta)^{\otimes n}\ket{+}^{\otimes n}\\
&\propto \Tilde{X}\bar{P}\ket{\theta}^{\otimes n}
\end{split}\end{equation}

where we used the fact that $XR_{\mathrm{Z}}(\theta) = R_{\mathrm{Z}}(-\theta)X$.  From the second last row to the last row, we explicitly expanded $\bigotimes_{\{m_i\}} (\cos\theta I + i \sin\theta Z_{m_i})$ into sum of terms with pure $Z$-type Pauli operators. These who commute with every stabilizer will  be either logical $I$ or $Z$ operators, but Lemma \ref{lemma:support} prevents them to accumulate into logical $Z$. These that don't commute with every stabilizer will get eliminated by the codespace projector $\bar{P}$ as we just showed in Equation \ref{eq: Z_type}. Notice that $\Tilde{X}$ must commute with the logical $X$ operator of $\mathcal{Q}$, the effect of $\Tilde{X}$ on the logical states after decoding can only be either a logical $X$ error if it anticommutes with the logical $Z$ operators or identity in the other case. Therefore, $\Tr_{\mathrm{A}}[D_\mathcal{Q}\bar{P}_{\bf{x}}(\ket{\theta}\bra{\theta})^{\otimes n}\bar{P}_{\bf{x}}D^\dagger_\mathcal{Q}]$ should be either $X\ket{\theta}\bra{\theta}X$ or $\ket{\theta}\bra{\theta}$.

Summarizing all up, when $\rho_{\mathrm{in}}$ is the ideal input state, the output state $\rho_{\mathrm{o}}$ is given by
\begin{equation}
    \rho_{\mathrm{o}}(\ket{\theta}\bra{\theta}) \propto p_{\bf{0}}\ket{\theta}\bra{\theta} + \sum_{\{g_{\mathrm{z}}\}} e^{-2\beta} p_{g_{\mathrm{z}}} X\ket{\theta}\bra{\theta}X + \sum_{\{g'_{\mathrm{z}}\}} e^{-2\beta} p_{g'_{\mathrm{z}}}\ket{\theta}\bra{\theta}  + O(e^{-4\beta}),
\end{equation}
where $\{g_{\mathrm{z}}\}(\{g'_{\mathrm{z}}\})$ is the set of $Z$-type generators whose subspace-transfer operators anti-commute (commute) with logical $Z$. $p_{\bf{0}} = \Tr[\bar{P}_{\bf{0}}(\ket{\theta}\bra{\theta})^{\otimes n}]$ and $p_{g_{\mathrm{z}}} = \Tr[\bar{P}_{g_{\mathrm{z}}}(\ket{\theta}\bra{\theta})^{\otimes n}]$ and $\bar{P}_{g_{\mathrm{z}}}$ is the subspace projector that is stabilized by every generator but antistabilized by $g_{\mathrm{z}}$.

Now we wanna find $\rho_{*}$ such that $\rho_{\mathrm{o}}(\rho_{*})=\rho_{*}$. As we only have infidelity contribution from $Z$-type generator measurement that could lead to logical $X$ noise, we may set $\rho_{*} = (1-\alpha_*)\ket{\theta}\bra{\theta} + \alpha_*X\ket{\theta}\bra{\theta}X$, and $\alpha_*$ should be small when $\beta$ is large. We can also consider the leading order of $\rho_{\mathrm{in}}$:
\begin{equation}
\rho_{\mathrm{in}}\approx (1-n\alpha_*)(\ket{\theta}\bra{\theta})^{\otimes n} + \alpha_* \sum^n_{i=1}X_i (\ket{\theta}\bra{\theta})^{\otimes n} X_i  .
\label{eq: rho_in_noise}
\end{equation}
Combine Equation \eqref{eq: rho_in_noise} and \eqref{eq: post_meas_state} we have:
\begin{equation}
\begin{split}
    \rho_{\mathrm{o}}(\rho_{*}) &\propto p^*_{\bf{0}}\ket{\theta}\bra{\theta} +  (1-n\alpha_*)\sum_{|\mathbf{x}|=1}e^{-2\beta}\Tr_{\mathrm{A}}[D_\mathcal{Q}\bar{P}_{\bf{x}} \ket{\theta}\bra{\theta} \bar{P}_{\bf{x}}D^\dagger_\mathcal{Q}] + \alpha_*\sum_{|\mathbf{x}|=1}e^{-2\beta}\Tr_{\mathrm{A}}[D_\mathcal{Q}\bar{P}_{\bf{x}} X_i\ket{\theta}\bra{\theta}X_i\bar{P}_{\bf{x}}D^\dagger_\mathcal{Q}]\\
    &\approx p^*_{\bf{0}}\ket{\theta}\bra{\theta} + (1-n\alpha_*)(\sum_{g_{\mathrm{z}}} e^{-2\beta} p_{g_{\mathrm{z}}} X\ket{\theta}\bra{\theta}X + \sum_{g'_{\mathrm{z}}} e^{-2\beta} p_{g'_{\mathrm{z}}}\ket{\theta}\bra{\theta}).
\end{split}
\end{equation}
As $\rho_{\mathrm{o}}(\rho_{*})=\rho_{*}$:
\begin{equation}
    \frac{\alpha_*}{1-\alpha_*} = \frac{(1-n\alpha_*)\sum_{g_{\mathrm{z}}} e^{-2\beta} p_{g_{\mathrm{z}}}}{p^*_{\bf{0}} + (1-n\alpha_*)\sum_{g'_{\mathrm{z}}} e^{-2\beta} p_{g'_{\mathrm{z}}} }.
\end{equation}
Define $A=\sum_{g_{\mathrm{z}}} p_{g_{\mathrm{z}}}$ and $B=\sum_{g'_{\mathrm{z}}} p_{g'_{\mathrm{z}}}$ for simplicity, from which we have
\begin{equation}
    \alpha_* \approx \dfrac{e^{-2\beta} A }{p^*_{\bf 0} + (n+1)Ae^{-2\beta} + Be^{-2\beta}} \propto e^{-2\beta}
\end{equation}
Therefore, the target state $\rho_{*}$ is at most under a Pauli $X$ noise with strength scales with $e^{-2\beta}$ if $A\neq0$.
\end{proof}

\begin{theorem}
     For non-trivial MSD protocols based on $[[n, k, d]]$ ($d\geq 2$)  CSS codes, the distillation convergence rate under finite weak measurement strength is linear. 
\end{theorem}

Proof key: Assume the input state to be all the noisy target unless one with epsilon deviation. The final term in Eq (51) would then be easy to prove to be non-zero.

\begin{proof}
Again, we may assume the Gaussian noise to clearly denote the noise order. For simplicity we define a linear quantum map from $n$-qubit space to $k$-qubit space 
\begin{equation}
    \mathcal{F}_\mathbf{x}(\cdot) = \Tr_{\mathrm{A}}[D_\mathcal{Q}\bar{P}_{\bf{x}} \cdot \bar{P}_{\bf{x}}D^\dagger_\mathcal{Q}].
\end{equation}
We define another map $\mathcal{D}$ for MSD protocols based on $\mathcal{Q}$ under measurement strength $\beta$.  Therefore, from Equation \eqref{eq: general_MSD_weak_form} we have:
\begin{equation}
    \rho_{\mathrm{o}}=\mathcal{D}(\rho^{\otimes n}_{\mathrm{i}}) = \sum_{\bf{x}}e^{-2|\bf{x}|\beta}F_{\mathbf{x}}(\rho^{\otimes n}_{\mathrm{i}})/\sum_{\bf{x}}e^{-2|\mathbf{x}|\beta}\Tr[\rho^{\otimes n}_{\mathrm{i}} P_{\mathbf{x}}] \propto\sum_{\bf{x}}e^{-2|\bf{x}|\beta}F_{\mathbf{x}}(\rho^{\otimes n}_{\mathrm{i}})
\end{equation}

Without loss of generality we consider $k=1$ first. Consider a fixed relatively large  $\beta\gg1$ with fixed point $\rho_{*}$, such that $\rho_{*} = \mathcal{D}(\rho_{*})$. Then we can always write $\rho_{*}$ in the following form:
\begin{equation}
    \rho_{*} = (1-m_{\mathrm{x}}-m_{\mathrm{y}})\ket{\theta}\bra{\theta} + m_{\mathrm{x}} X\ket{\theta}\bra{\theta}X + m_{\mathrm{y}} Y\ket{\theta}\bra{\theta}Y,
\end{equation}
where $m_{\mathrm{x}}\propto e^{-2q_{\mathrm{x}}\beta}$ and $m_{\mathrm{y}}\propto e^{-2q_{\mathrm{y}}\beta}$. We assume the purity of $\rho_{*}$ to be $\delta$. Without loss of generality we assume $q_{\mathrm{x}}\leq q_{\mathrm{y}}$, so $q=q_{\mathrm{x}}$.
 Now consider an input state $\rho_{\epsilon}$ such that $||\rho_{*} - \rho_{\epsilon} || \propto \epsilon$, we may assume
 \begin{equation}
     \rho_{\epsilon} = (1-\epsilon)\rho_{*} + \epsilon (u_{\mathrm{x}}X\rho_{*}X+u_{\mathrm{y}}Y\rho_{*}Y+u_{\mathrm{z}} Z\rho_{*} Z)  
 \end{equation}
 
Define a channel for $n$-qubit states $\mathcal{N}(\rho) = \frac{1}{3n}\sum_{E\in\mathcal{P}_1}u_E E\rho E $ where $\mathcal{P}_1$ is the set for all non-trivial single-weight Pauli operators on $n$ qubits.  Therefore, $\rho^{\otimes n}_\epsilon \approx (1-n\epsilon)\rho^{\otimes n}_* + n\epsilon \mathcal{N}(\rho_{*}^{\otimes n})$. Define $\tilde{\epsilon}=n\epsilon$:
\begin{equation}
    \mathcal{D}(\rho^{\otimes n}_\epsilon) = \dfrac{(1-\tilde{\epsilon})\sum_{\bf{x}}e^{-2|\bf{x}|\beta}F_{\mathbf{x}}(\rho_{*}^{\otimes n}) +  \tilde{\epsilon}  \sum_{\bf{x}}e^{-2|\bf{x}|\beta}F_{\mathbf{x}}(\mathcal{N}(\rho^{\otimes n}_*))}{ (1-\tilde{\epsilon})\sum_{\bf{x}}e^{-2|\mathbf{x}|\beta}\Tr[\rho^{\otimes n}_*P_{\bf{x}}] + \tilde{\epsilon} \sum_{\bf{x}}e^{-2|\bf{x}|\beta}\Tr[\mathcal{N}(\rho_{*}^{\otimes n})P_{\bf{x}}]}
\end{equation}

\begin{equation}
    \begin{split}
        ||\mathcal{D}(\rho^{\otimes n}_\epsilon) - \mathcal{D}(\rho^{\otimes n}_*) ||  &=||\dfrac{(1-\tilde{\epsilon})\sum_{\bf{x}}e^{-2|\bf{x}|\beta}F_{\mathbf{x}}(\rho_{*}^{\otimes n}) +  \tilde{\epsilon}  \sum_{\bf{x}}e^{-2|\bf{x}|\beta}F_{\mathbf{x}}(\mathcal{N}(\rho^{\otimes n}_*))}{ (1-\tilde{\epsilon})\sum_{\bf{x}}e^{-2|\mathbf{x}|\beta}\Tr[\rho^{\otimes n}_*P_{\bf{x}}] + \tilde{\epsilon} \sum_{\bf{x}}e^{-2|\bf{x}|\beta}\Tr[\mathcal{N}(\rho_{*}^{\otimes n})P_{\bf{x}}]} - \dfrac{\sum_{\bf{x}}e^{-2|\bf{x}|\beta}F_{\mathbf{x}}(\rho^{\otimes n}_*)}{\sum_{\bf{x}}e^{-2|\mathbf{x}|\beta}\Tr[\rho^{\otimes n}_* P_{\mathbf{x}}]}|| \\
        &\propto ||\dfrac{(1-\tilde{\epsilon})\sum_{\bf{x}}e^{-2|\bf{x}|\beta}F_{\mathbf{x}}(\rho_{*}^{\otimes n}) +  \tilde{\epsilon}  \sum_{\bf{x}}e^{-2|\bf{x}|\beta}F_{\mathbf{x}}(\mathcal{N}(\rho^{\otimes n}_*))}{ (1-\tilde{\epsilon}) + \tilde{\epsilon} \sum_{\bf{x}}e^{-2|\bf{x}|\beta}\Tr[\mathcal{N}(\rho_{*}^{\otimes n})P_{\bf{x}}]/\sum_{\bf{x}}e^{-2|\mathbf{x}|\beta}\Tr[\rho^{\otimes n}_* P_{\mathbf{x}}]}  - \sum_{\bf{x}}e^{-2|\bf{x}|\beta}F_{\mathbf{x}}(\rho^{\otimes n}_*) || \\
        &= \tilde{\epsilon}||  - \dfrac{\sum_{\bf{x}}e^{-2|\bf{x}|\beta}\Tr[\mathcal{N}(\rho_{*}^{\otimes n})P_{\bf{x}}]}{\sum_{\bf{x}}e^{-2|\mathbf{x}|\beta}\Tr[\rho^{\otimes n}_* P_{\mathbf{x}}]}\sum_{\bf{x}}e^{-2|\bf{x}|\beta}F_{\mathbf{x}}(\rho_{*}^{\otimes n}) +   \sum_{\bf{x}}e^{-2|\bf{x}|\beta}F_{\mathbf{x}}(\mathcal{N}(\rho^{\otimes n}_*)) || + O(\tilde{\epsilon}^2)\\
        &\propto\tilde{\epsilon} ||\mathcal{D}(\rho_{*}^{\otimes n}) - \mathcal{D}(\mathcal{N}(\rho_{*}^{\otimes n})) || \quad (\text{if} \quad  \sum_{\bf{x}}e^{-2|\bf{x}|\beta}\Tr[\mathcal{N}(\rho_{*}^{\otimes n})P_{\bf{x}}]\neq 0)\\
        &= \tilde{\epsilon} ||\rho_{*} - \frac{\sum_{\bf{x}}\sum_{E\in\mathcal{P}_1}u_E e^{-2|\bf{x}|\beta}F_{\mathbf{x}}(E\rho_{*}^{\otimes n}E)}{\sum_{\bf{x}}\sum_{E\in\mathcal{P}_1}u_E e^{-2|\mathbf{x}|\beta}\Tr[\rho_{*}^{\otimes n} EP_{\mathbf{x}}E]} || \\
        & \propto \tilde{\epsilon} ||\sum_{\bf{x}}\sum_{E\in\mathcal{P}_1}u_E e^{-2|\bf{x}|\beta}[\rho_{*} * \Tr[\rho_{*}^{\otimes n} EP_{\mathbf{x}}E] - F_{\mathbf{x}}(E\rho_{*}^{\otimes n}E)] ||
    \end{split}
    \label{eq: final_diff}
\end{equation}

Now we prove by contradiction: If the convergence rate is beyond linear order, the prefactor in the last row in \eqref{eq: final_diff} must be zero. Notice that the convergence doesn't depend on specific direction, the prefactor must also be zero for all potential combination of $u_E$ in the channel $\mathcal{N}$. Therefore, the prefactor should also be zero for any single Pauli error $E\in\mathcal{P}_1$. Besides, as we
are in the large $\beta$ regime, We only require the coefficient for the same order of $e^{-2\beta}$ to be zero. Based on these reasons, we must require
\begin{equation}
    || \sum_{|\mathbf{x}|=j}[\rho_{*} * \Tr[\rho_{*}^{\otimes n} EP_{\mathbf{x}}E] - F_{\mathbf{x}}(E\rho_{*}^{\otimes n}E)] || = 0 , \quad \forall \  E\in \mathcal{P}_1, \ \ 0\leq j \leq \bar{n}.
\end{equation}

Notice that $F_{\mathbf{x}}(E\rho_{*}^{\otimes n}E) = C_EF_{\mathbf{y}(\mathbf{x},E)}(\rho_{*}^{\otimes n}) C_E $, where $C_E$ is a single-qubit Pauli associated with $E_i$:
\begin{equation}
    C_E = \begin{cases}
        X \quad \text{if} \quad [E, \bar{Z}]\neq0, [E, \bar{X}]=0 \\
        Z \quad \text{if} \quad [E, \bar{Z}]=0, [E, \bar{X}]\neq0\\
        Y \quad \text{if} \quad [E, \bar{Z}]\neq0, [E, \bar{X}]\neq0\\
        I \quad \text{other cases}
    \end{cases}
\end{equation}
and $\mathbf{y}$ is another binary vector associated with $\mathbf{x}$ and $E$ which strictly follows $\mathbf{y}(\mathbf{x},E_i)\neq \mathbf{x}$ such that
\begin{equation}
    (\mathbf{y}(\mathbf{x}, E) \oplus \mathbf{x})_k = \begin{cases}
        0 \quad \text{if} \quad [E,g_k] = 0\\
        1 \quad \text{if} \quad [E,g_k] \neq 0
    \end{cases} .
\end{equation}
Therefore, the Eq (40) would be
\begin{equation}
        || \sum_{|\mathbf{x}|=j}[\rho_{*} * \Tr[\rho_{*}^{\otimes n} P_{\mathbf{y}(\mathbf{x},E)}] - C_E F_{\mathbf{y}(\mathbf{x},E)}(\rho_{*}^{\otimes n})C_E]  || = 0 , \quad \forall \  E\in \mathcal{P}_1, \ \ 0\leq j \leq \bar{n}.
\end{equation}

Only consider $j=0$: the above equation means the distillation protocol with ideal measurements should distill $\rho_{*}$ at at least quadratic efficiency. However, as the distillation protocol already distills into the pure magic state $\ket{\theta}$ and $\rho_{*}$ is a close mixed state, this is not possible. Therefore, even the coefficient for $j=0$ won't be zero. As a result, the distillation under imperfect measurement can only exhibit linear convergence rate.
\end{proof}

\begin{theorem}
    For MSD protocols based on $[[n, k, d]](d\geq 2)$ CSS codes with transversal non-Clifford gates, we can choose the stabilizer generators in standard form, such that the asymptotically distilled states are robust to weak measurement up to order $d$.
    \label{th:standard:supp}
\end{theorem}
\begin{proof}
Let's consider the effect of $\bar{P}_{\mathbf{x}}$ with $|\mathbf{x}|\leq d-1$. Then there exist some subspace-transfer operators $C_{\mathbf{x}}$ such that  $\bar{P}_{\mathbf{x}} = C_{\mathbf{x}}\bar{P}_{\bf{0}}C_{\mathbf{x}}$ . Importantly, $C_{\mathbf{x}}=\prod_{\{i: x_i\neq0\}}C_i$ is a product of subspace-transfer operators for $|\mathbf{x}|=1$. From Lemma \ref{lem: standard_corr} we see all $C_i$ is single-weight and have different support. Therefore, $C_\mathbf{x}$ has weight smaller than $d$, and $\bar{P}_{\mathbf{0}} C_{\mathbf{x}}\ket{\bar\psi} \propto \ket{\bar \psi}$ for any logical state $\ket{\bar \psi}$. Without loss of generality we might assume $C_{\bf{x}}$ on the first $m$ qubits are Pauli $X$. Let's then consider the effect of $P_{\mathbf{x}}$ on $\ket{\theta}^{\otimes n}$:
\begin{equation}
\begin{split}
   \bar{ P}_{\mathbf{x}}\ket{\theta}^{\otimes n} &= C_{\bf{x}} \bar{P}_{\bf{0}} C_{\bf{x}} R_{\mathrm{Z}}(\theta)^{\otimes n} \ket{+}^{\otimes n}\\
   & = C_{\bf{x}} \bar{P}_{\bf{0}} R_{\mathrm{Z}}(\theta)^{\otimes n}\bigotimes^m_{j=1} R_{\mathrm{Z}, j}(-2\theta) C_{\bf{x}}  \ket{+}^{\otimes n}\\
    & = C_{\bf{x}}   R_{\mathrm{Z}}(\theta)^{\otimes n} \bar{P}_{\bf{0}}(\bigotimes^m_{j=1} R_{\mathrm{Z}, j}(-2\theta) )\bar{C} \ket{+}^{\otimes n}\\
    & \propto C_{\bf{x}}  R_{\mathrm{Z}}(\theta)^{\otimes n} \bar{P}_{\bf{0}} \ket{+}^{\otimes n} = C_{\bf{x}}   \bar{P}_{\bf{0}} \ket{\theta}^{\otimes n}.\\
\end{split}
\end{equation}

From Lemma \ref{lem: standard_corr}, we also know that $C_{\bf{x}}$ must commute with all logical operators. Therefore, it will act as if a logical identity and $C_{\bf{x}}   \bar{P}_{\bf{0}} \ket{\theta}^{\otimes n}$ will give exactly the same state as $\bar{P}_{\bf{0}} \ket{\theta}^{\otimes n}$ after decoding.

Combining \eqref{eq: general_MSD_weak_form}, we have 
\begin{equation}
\begin{split}
        \rho_{\mathrm{o}}(\ket{\theta}\bra{\theta}) \propto p_{0}\ket{\theta}\bra{\theta} + \sum_{|\mathbf{x}|\geq d}e^{-2|\bf{x}|\beta}\Tr_{\mathrm{A}}[D_\mathcal{Q}\bar{P}_{\bf{x}} \rho_{\mathrm{in}} \bar{P}_{\bf{x}}D^\dagger_\mathcal{Q}]
\end{split}
\end{equation}
Therefore, we only have infidelity contribution with lead order $e^{-2d\beta}$.
\end{proof}

\section{Supplementary Note 6: Eigenvalue of Jacobian matrix at fixed point}
\label{app:jacobian}

\begin{figure}[H]
    \centering
    \includegraphics[width=0.6\textwidth]{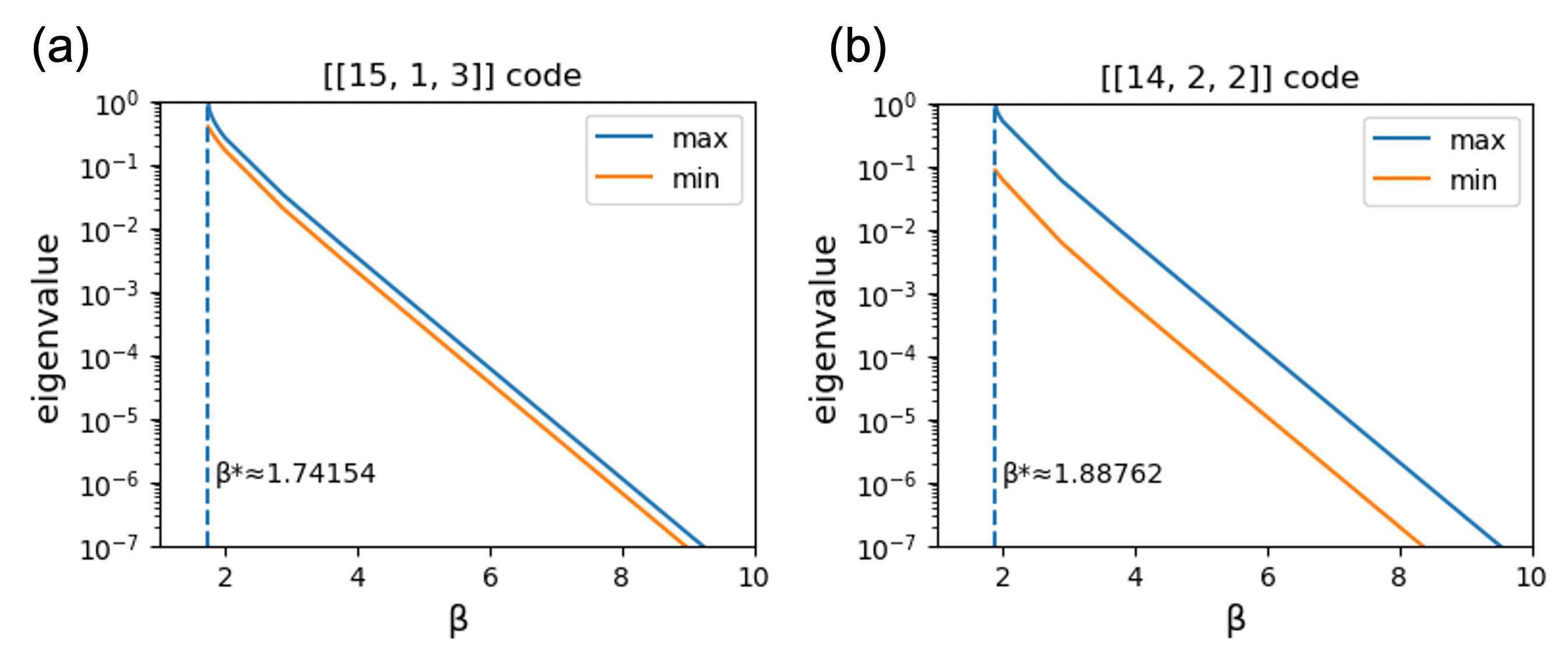}
    \caption{\textbf{The Jacobian eigenvalue identifies the non-trivial fixed-point stability of the two example MSD protocols.} Results are shown for (a) the $[[15, 1, 3]]$ code and (b) the $[[14, 2, 2]]$ code, using the canonical generator sets.}
    \label{fig:jacobian}
\end{figure}

As shown in Figure \ref{fig:jacobian}, we calculate the eigenvalue of the Jacobian matrix at the non-trivial fixed points for the simulated MSD protocols. The largest eigenvalue increases when measurement strength decreases. When the largest eigenalue reaches to 1, the fixed point is no longer stable and will disappear, which corresponds to the threshold of measurement strength.
\FloatBarrier
\section{Supplementary Note 7: Other choice of stabilizer generators}

\begin{figure}[H]
    \centering
    \includegraphics[width=\textwidth]{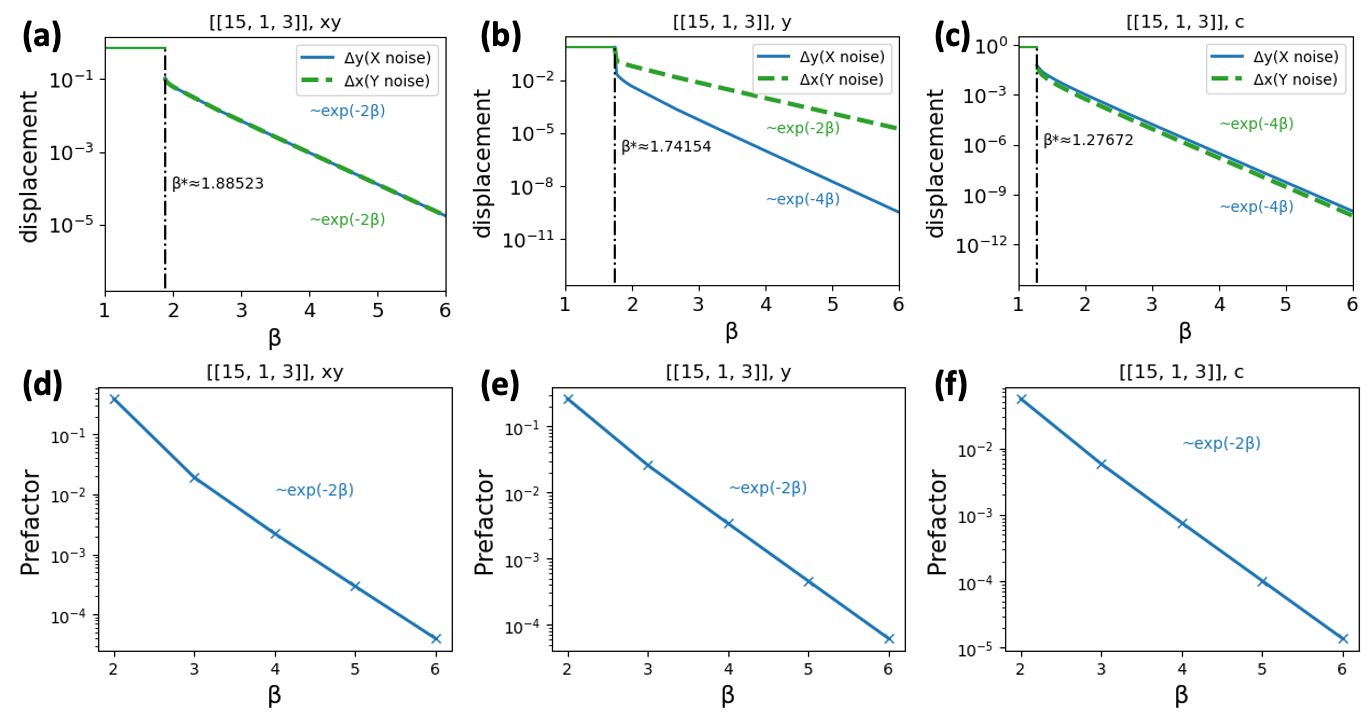}
    \caption{\textbf{Different stabilizer-generator choices alter the performance of the $[[15, 1, 3]]$ protocol.} Panels (a) and (d) show the output-state displacement and corresponding prefactor for the $XY$-symmetric choice. Panels (b) and (e) show the same quantities after replacing all $X$-type generators with $Y$-type generators. Panels (c) and (f) show the results for the modified standard form.}
    \label{fig:other_choice}
\end{figure}

In this section, we again provide some examples that show that we can actually engineer the biased Pauli noise by switch the type of stabilizer generators. For the $[[15, 1, 3]]$ code, $H_{\mathrm{X}} \subset H_{\mathrm{Z}}$ (see Sec \ref{app: code_description}). Therefore, we can modify all weight-8 $Z$-type generators to $Y$-type and with the codespace intact. In this generator choice, both the $X$-type and $Y$-type noise are first-order as shown in Figure \ref{fig:other_choice}(a,d).  We can also replace all the $X$-type generators with $Y$-type generators that share the same support and the codespace is still exactly the same, which make us able to engineer the leading order biased noise in the output state from logical Pauli $X$ to Pauli $Y$ compared to the canonical generator choice as shown in Figure \ref{fig:other_choice}(b,e). Notably, the two examples don't conflict with our Theorem 1 as now the code is no longer strictly CSS since $Y$-type generators are introduced. Further, we might consider a modified standard-form generator choice, such that $H_{\mathrm{Z}}$ is formed as:
\begin{equation}
    H_{\mathrm{Z}}=\begin{bmatrix}
        0 & 0 & 0 \\
        D' & I' & E'
    \end{bmatrix}
\end{equation}
where $I'$ is an upper triangular matrix with all upper element being one. This $H_{\mathrm{Z}}$ can be obtained by performing the reverse of gaussian elimination. This generator choice will give us performance as shown in Figure \ref{fig:other_choice}(c,f).

\FloatBarrier

\clearpage
\section*{References}
\bibliography{MSD_weak_meas}

@misc{zhao2023latticegaugetheorytopological,
      title={Lattice gauge theory and topological quantum error correction with quantum deviations in the state preparation and error detection}, 
      author={Yuanchen Zhao and Dong E. Liu},
      year={2023},
      eprint={2301.12859},
      archivePrefix={arXiv},
      primaryClass={quant-ph},
      url={https://arxiv.org/abs/2301.12859}, 
}

@article{kubicaUniversalTransversalGates2015,
  title = {Universal Transversal Gates with Color Codes - a Simplified Approach},
  author = {Kubica, Aleksander and Beverland, Michael E.},
  year = {2015},
  month = mar,
  journal = {Physical Review A},
  volume = {91},
  number = {3},
  eprint = {1410.0069},
  primaryclass = {quant-ph},
  pages = {032330},
  issn = {1050-2947, 1094-1622},
  doi = {10.1103/PhysRevA.91.032330},
  urldate = {2024-12-18},
  archiveprefix = {arXiv}
}

@article{maHighfidelityGatesMidcircuit2023a,
  title = {High-Fidelity Gates and Mid-Circuit Erasure Conversion in an Atomic Qubit},
  author = {Ma, Shuo and Liu, Genyue and Peng, Pai and Zhang, Bichen and Jandura, Sven and Claes, Jahan and Burgers, Alex P. and Pupillo, Guido and Puri, Shruti and Thompson, Jeff D.},
  year = {2023},
  month = oct,
  journal = {Nature},
  volume = {622},
  number = {7982},
  pages = {279--284},
  issn = {0028-0836, 1476-4687},
  doi = {10.1038/s41586-023-06438-1},
  urldate = {2023-11-06},
  langid = {english}
}

@article{krishnaFaulttolerantGatesHypergraph2021a,
  title = {Fault-Tolerant Gates on Hypergraph Product Codes},
  author = {Krishna, Anirudh and Poulin, David},
  year = {2021},
  month = feb,
  journal = {Physical Review X},
  volume = {11},
  number = {1},
  eprint = {1909.07424},
  primaryclass = {quant-ph},
  pages = {011023},
  issn = {2160-3308},
  doi = {10.1103/PhysRevX.11.011023},
  urldate = {2024-12-07},
  archiveprefix = {arXiv}
}

@misc{landahlFaulttolerantQuantumComputing2011a,
  title = {Fault-Tolerant Quantum Computing with Color Codes},
  author = {Landahl, Andrew J. and Anderson, Jonas T. and Rice, Patrick R.},
  year = {2011},
  month = aug,
  number = {arXiv:1108.5738},
  eprint = {1108.5738},
  primaryclass = {quant-ph},
  publisher = {arXiv},
  doi = {10.48550/arXiv.1108.5738},
  urldate = {2024-12-07},
  archiveprefix = {arXiv}
}

@article{moussaTransversalCliffordGates2016,
  title = {Transversal {{Clifford}} Gates on Folded Surface Codes},
  author = {Moussa, Jonathan E.},
  year = {2016},
  month = oct,
  journal = {Physical Review A},
  volume = {94},
  number = {4},
  eprint = {1603.02286},
  primaryclass = {quant-ph},
  pages = {042316},
  issn = {2469-9926, 2469-9934},
  doi = {10.1103/PhysRevA.94.042316},
  urldate = {2024-12-07},
  archiveprefix = {arXiv}
}

@misc{paetznickFaulttolerantAncillaPreparation2013,
  title = {Fault-Tolerant Ancilla Preparation and Noise Threshold Lower Bounds for the 23-Qubit {{Golay}} Code},
  author = {Paetznick, Adam and Reichardt, Ben W.},
  year = {2013},
  month = apr,
  number = {arXiv:1106.2190},
  eprint = {1106.2190},
  primaryclass = {quant-ph},
  publisher = {arXiv},
  doi = {10.48550/arXiv.1106.2190},
  urldate = {2024-12-07},
  archiveprefix = {arXiv}
}

@misc{zhengMagicStateDistillation2024,
  title = {From {{Magic State Distillation}} to {{Dynamical Systems}}},
  author = {Zheng, Yunzhe and Liu, Dong E.},
  year = {2024},
  month = dec,
  number = {arXiv:2412.04402},
  eprint = {2412.04402},
  publisher = {arXiv},
  doi = {10.48550/arXiv.2412.04402},
  urldate = {2024-12-06},
  archiveprefix = {arXiv},
  copyright = {All rights reserved}
}

@misc{willsConstantOverheadMagicState2024,
  title = {Constant-{{Overhead Magic State Distillation}}},
  author = {Wills, Adam and Hsieh, Min-Hsiu and Yamasaki, Hayata},
  year = {2024},
  month = aug,
  number = {arXiv:2408.07764},
  eprint = {2408.07764},
  publisher = {arXiv},
  doi = {10.48550/arXiv.2408.07764},
  urldate = {2024-10-23},
  archiveprefix = {arXiv}
}

@article{shabaniContinuousMeasurementNonMarkovian2014,
  title = {Continuous {{Measurement}} of a {{Non-Markovian Open Quantum System}}},
  author = {Shabani, A. and Roden, J. and Whaley, K. B.},
  year = {2014},
  month = mar,
  journal = {Physical Review Letters},
  volume = {112},
  number = {11},
  pages = {113601},
  publisher = {American Physical Society},
  doi = {10.1103/PhysRevLett.112.113601},
  urldate = {2024-10-23}
}

@article{lloydQuantumFeedbackWeak2000,
  title = {Quantum Feedback with Weak Measurements},
  author = {Lloyd, Seth and Slotine, Jean-Jacques E.},
  year = {2000},
  month = jun,
  journal = {Physical Review A},
  volume = {62},
  number = {1},
  pages = {012307},
  issn = {1050-2947, 1094-1622},
  doi = {10.1103/PhysRevA.62.012307},
  urldate = {2024-10-23},
  copyright = {http://link.aps.org/licenses/aps-default-license},
  langid = {english}
}

@misc{andersonClassificationTransversalGates2014,
  title = {Classification of Transversal Gates in Qubit Stabilizer Codes},
  author = {Anderson, Jonas T. and {Jochym-O'Connor}, Tomas},
  year = {2014},
  month = sep,
  number = {arXiv:1409.8320},
  eprint = {1409.8320},
  primaryclass = {math-ph, physics:quant-ph},
  publisher = {arXiv},
  urldate = {2024-03-18},
  archiveprefix = {arXiv}
}

@article{bravyiMagicstateDistillationLow2012,
  title = {Magic-State Distillation with Low Overhead},
  author = {Bravyi, Sergey and Haah, Jeongwan},
  year = {2012},
  month = nov,
  journal = {Physical Review A},
  volume = {86},
  number = {5},
  pages = {052329},
  issn = {1050-2947, 1094-1622},
  doi = {10.1103/PhysRevA.86.052329},
  urldate = {2024-02-05},
  langid = {english}
}

@article{bravyiUniversalQuantumComputation2005,
  title = {Universal Quantum Computation with Ideal {{Clifford}} Gates and Noisy Ancillas},
  author = {Bravyi, Sergey and Kitaev, Alexei},
  year = {2005},
  month = feb,
  journal = {Physical Review A},
  volume = {71},
  number = {2},
  pages = {022316},
  issn = {1050-2947, 1094-1622},
  doi = {10.1103/PhysRevA.71.022316},
  urldate = {2024-01-05},
  langid = {english}
}

@misc{campbellStructureProtocolsMagic2009,
  title = {On the {{Structure}} of {{Protocols}} for {{Magic State Distillation}}},
  author = {Campbell, Earl T. and Browne, Dan E.},
  year = {2009},
  month = aug,
  number = {arXiv:0908.0838},
  eprint = {0908.0838},
  primaryclass = {quant-ph},
  publisher = {arXiv},
  urldate = {2024-01-25},
  archiveprefix = {arXiv}
}

@article{clerkIntroductionQuantumNoise2010,
  title = {Introduction to Quantum Noise, Measurement, and Amplification},
  author = {Clerk, A. A. and Devoret, M. H. and Girvin, S. M. and Marquardt, Florian and Schoelkopf, R. J.},
  year = {2010},
  month = apr,
  journal = {Reviews of Modern Physics},
  volume = {82},
  number = {2},
  pages = {1155--1208},
  publisher = {American Physical Society},
  doi = {10.1103/RevModPhys.82.1155},
  urldate = {2024-10-06}
}

@article{durQuantumRepeatersBased1999,
  title = {Quantum Repeaters Based on Entanglement Purification},
  author = {D{\"u}r, W. and Briegel, H.-J. and Cirac, J. I. and Zoller, P.},
  year = {1999},
  month = jan,
  journal = {Physical Review A},
  volume = {59},
  number = {1},
  pages = {169--181},
  issn = {1050-2947, 1094-1622},
  doi = {10.1103/PhysRevA.59.169},
  urldate = {2024-09-22},
  copyright = {http://link.aps.org/licenses/aps-default-license},
  langid = {english}
}

@article{eastinRestrictionsTransversalEncoded2009,
  title = {Restrictions on {{Transversal Encoded Quantum Gate Sets}}},
  author = {Eastin, Bryan and Knill, Emanuel},
  year = {2009},
  month = mar,
  journal = {Physical Review Letters},
  volume = {102},
  number = {11},
  pages = {110502},
  issn = {0031-9007, 1079-7114},
  doi = {10.1103/PhysRevLett.102.110502},
  urldate = {2024-03-19},
  langid = {english}
}

@article{hastingsDistillationSublogarithmicOverhead2018,
  title = {Distillation with {{Sublogarithmic Overhead}}},
  author = {Hastings, Matthew B. and Haah, Jeongwan},
  year = {2018},
  month = jan,
  journal = {Physical Review Letters},
  volume = {120},
  number = {5},
  pages = {050504},
  publisher = {American Physical Society},
  doi = {10.1103/PhysRevLett.120.050504},
  urldate = {2024-03-06}
}

@article{jonesMultilevelDistillationMagic2013a,
  title = {Multilevel Distillation of Magic States for Quantum Computing},
  author = {Jones, Cody},
  year = {2013},
  month = apr,
  journal = {Physical Review A},
  volume = {87},
  number = {4},
  pages = {042305},
  issn = {1050-2947, 1094-1622},
  doi = {10.1103/PhysRevA.87.042305},
  urldate = {2024-03-08},
  langid = {english}
}

@phdthesis{brooksQuantumErrorCorrection2013,
  title = {Quantum Error Correction with Biased Noise},
  author = {Brooks, Peter},
  year = {2013},
  school = {California Institute of Technology},
  note = {Chapter 4: ``Magic State Distillation with Noisy Clifford Gates''}
}

@article{jochymoconnorRobustnessMagicState2013,
  title = {The Robustness of Magic State Distillation against Errors in Clifford Gates},
  author = {{Jochym-O'Connor}, Tomas and Yu, Yafei and Helou, Bassam and Laflamme, Raymond},
  year = {2013},
  journal = {Quantum Information \& Computation},
  volume = {13},
  number = {5--6},
  pages = {361--378}
}

@article{krishnaLowOverheadMagic2019,
  title = {Towards {{Low Overhead Magic State Distillation}}},
  author = {Krishna, Anirudh and Tillich, Jean-Pierre},
  year = {2019},
  month = aug,
  journal = {Physical Review Letters},
  volume = {123},
  number = {7},
  pages = {070507},
  publisher = {American Physical Society},
  doi = {10.1103/PhysRevLett.123.070507},
  urldate = {2024-03-06}
}

@article{litinskiMagicStateDistillation2019,
  title = {Magic {{State Distillation}}: {{Not}} as {{Costly}} as {{You Think}}},
  shorttitle = {Magic {{State Distillation}}},
  author = {Litinski, Daniel},
  year = {2019},
  month = dec,
  journal = {Quantum},
  volume = {3},
  eprint = {1905.06903},
  primaryclass = {quant-ph},
  pages = {205},
  issn = {2521-327X},
  doi = {10.22331/q-2019-12-02-205},
  urldate = {2024-01-26},
  archiveprefix = {arXiv},
  langid = {english}
}

@misc{meierMagicstateDistillationFourqubit2012,
  title = {Magic-State Distillation with the Four-Qubit Code},
  author = {Meier, Adam M. and Eastin, Bryan and Knill, Emanuel},
  year = {2012},
  month = apr,
  number = {arXiv:1204.4221},
  eprint = {1204.4221},
  primaryclass = {quant-ph},
  publisher = {arXiv},
  urldate = {2023-10-13},
  archiveprefix = {arXiv}
}

@book{nielsenQuantumComputationQuantum2010,
  title = {Quantum Computation and Quantum Information},
  author = {Nielsen, Michael A and Chuang, Isaac L},
  year = {2010},
  publisher = {Cambridge university press}
}

@article{reichardtImprovedMagicStates2005,
  title = {Improved Magic States Distillation for Quantum Universality},
  author = {Reichardt, Ben W.},
  year = {2005},
  month = aug,
  journal = {Quantum Information Processing},
  volume = {4},
  number = {3},
  eprint = {quant-ph/0411036},
  pages = {251--264},
  issn = {1570-0755, 1573-1332},
  doi = {10.1007/s11128-005-7654-8},
  urldate = {2024-06-19},
  archiveprefix = {arXiv},
  langid = {english}
}

@article{reichardtQuantumUniversalityMagic2005,
  title = {Quantum {{Universality}} from {{Magic States Distillation Applied}} to {{CSS Codes}}},
  author = {Reichardt, Ben W.},
  year = {2005},
  month = aug,
  journal = {Quantum Information Processing},
  volume = {4},
  number = {3},
  pages = {251--264},
  issn = {1573-1332},
  doi = {10.1007/s11128-005-7654-8},
  urldate = {2024-03-08},
  langid = {english}
}

@misc{reichardtQuantumUniversalityState2009,
  title = {Quantum Universality by State Distillation},
  author = {Reichardt, Ben W.},
  year = {2009},
  month = jul,
  number = {arXiv:quant-ph/0608085},
  eprint = {quant-ph/0608085},
  publisher = {arXiv},
  urldate = {2024-03-08},
  archiveprefix = {arXiv}
}

@misc{zengTransversalityUniversalityAdditive2007,
  title = {Transversality versus {{Universality}} for {{Additive Quantum Codes}}},
  author = {Zeng, Bei and Cross, Andrew and Chuang, Isaac L.},
  year = {2007},
  month = sep,
  number = {arXiv:0706.1382},
  eprint = {0706.1382},
  primaryclass = {quant-ph},
  publisher = {arXiv},
  urldate = {2024-04-14},
  archiveprefix = {arXiv}
}

@misc{gottesmanStabilizerCodesQuantum1997,
  title = {Stabilizer {{Codes}} and {{Quantum Error Correction}}},
  author = {Gottesman, Daniel},
  year = {1997},
  month = may,
  number = {arXiv:quant-ph/9705052},
  eprint = {quant-ph/9705052},
  publisher = {arXiv},
  urldate = {2023-06-12},
  archiveprefix = {arXiv}
}

@article{zhuNishimorisCatStable2023,
  title = {Nishimori's {{Cat}}: {{Stable Long-Range Entanglement}} from {{Finite-Depth Unitaries}} and {{Weak Measurements}}},
  shorttitle = {Nishimori's {{Cat}}},
  author = {Zhu, Guo-Yi and Tantivasadakarn, Nathanan and Vishwanath, Ashvin and Trebst, Simon and Verresen, Ruben},
  year = {2023},
  month = nov,
  journal = {Physical Review Letters},
  volume = {131},
  number = {20},
  pages = {200201},
  publisher = {American Physical Society},
  doi = {10.1103/PhysRevLett.131.200201},
  urldate = {2024-10-06}
}

@article{PRXQuantum.4.030317,
  title = {Quantum Criticality Under Decoherence or Weak Measurement},
  author = {Lee, Jong Yeon and Jian, Chao-Ming and Xu, Cenke},
  journal = {PRX Quantum},
  volume = {4},
  issue = {3},
  pages = {030317},
  numpages = {20},
  year = {2023},
  month = {Aug},
  publisher = {American Physical Society},
  doi = {10.1103/PRXQuantum.4.030317},
  url = {https://link.aps.org/doi/10.1103/PRXQuantum.4.030317}
}

@article{lee2024low,
  title={Low-overhead magic state distillation with color codes},
  author={Lee, Seok-Hyung and Thomsen, Felix and Fazio, Nicholas and Brown, Benjamin J and Bartlett, Stephen D},
  journal={arXiv preprint arXiv:2409.07707},
  year={2024}
}

@article{PhysRevB.110.064301,
  title = {Critical properties of weak measurement induced phase transitions in random quantum circuits},
  author = {Aziz, Kemal and Chakraborty, Ahana and Pixley, J. H.},
  journal = {Phys. Rev. B},
  volume = {110},
  issue = {6},
  pages = {064301},
  numpages = {15},
  year = {2024},
  month = {Aug},
  publisher = {American Physical Society},
  doi = {10.1103/PhysRevB.110.064301},
  url = {https://link.aps.org/doi/10.1103/PhysRevB.110.064301}
}

@article{haah2018codes,
  title={Codes and protocols for distilling $ t $, controlled-$ s $, and toffoli gates},
  author={Haah, Jeongwan and Hastings, Matthew B},
  journal={Quantum},
  volume={2},
  pages={71},
  year={2018},
  publisher={Verein zur F{\"o}rderung des Open Access Publizierens in den Quantenwissenschaften}
}

@misc{rodriguez_experimental_2024,
    title = {Experimental {Demonstration} of {Logical} {Magic} {State} {Distillation}},
    url = {http://arxiv.org/abs/2412.15165},
    doi = {10.48550/arXiv.2412.15165},
    urldate = {2025-01-21},
    publisher = {arXiv},
    author = {Rodriguez, Pedro Sales and Robinson, John M. and Jepsen, Paul Niklas and He, Zhiyang and Duckering, Casey and Zhao, Chen and Wu, Kai-Hsin and Campo, Joseph and Bagnall, Kevin and Kwon, Minho and Karolyshyn, Thomas and Weinberg, Phillip and Cain, Madelyn and Evered, Simon J. and Geim, Alexandra A. and Kalinowski, Marcin and Li, Sophie H. and Manovitz, Tom and Amato-Grill, Jesse and Basham, James I. and Bernstein, Liane and Braverman, Boris and Bylinskii, Alexei and Choukri, Adam and DeAngelo, Robert and Fang, Fang and Fieweger, Connor and Frederick, Paige and Haines, David and Hamdan, Majd and Hammett, Julian and Hsu, Ning and Hu, Ming-Guang and Huber, Florian and Jia, Ningyuan and Kedar, Dhruv and Kornjača, Milan and Liu, Fangli and Long, John and Lopatin, Jonathan and Lopes, Pedro L. S. and Luo, Xiu-Zhe and Macrì, Tommaso and Marković, Ognjen and Martínez-Martínez, Luis A. and Meng, Xianmei and Ostermann, Stefan and Ostroumov, Evgeny and Paquette, David and Qiang, Zexuan and Shofman, Vadim and Singh, Anshuman and Singh, Manuj and Sinha, Nandan and Thoreen, Henry and Wan, Noel and Wang, Yiping and Waxman-Lenz, Daniel and Wong, Tak and Wurtz, Jonathan and Zhdanov, Andrii and Zheng, Laurent and Greiner, Markus and Keesling, Alexander and Gemelke, Nathan and Vuletić, Vladan and Kitagawa, Takuya and Wang, Sheng-Tao and Bluvstein, Dolev and Lukin, Mikhail D. and Lukin, Alexander and Zhou, Hengyun and Cantú, Sergio H.},
    month = dec,
    year = {2024},
    note = {arXiv:2412.15165 [quant-ph]},
}

@misc{brown_advances_2023,
    title = {Advances in compilation for quantum hardware -- {A} demonstration of magic state distillation and repeat-until-success protocols},
    url = {http://arxiv.org/abs/2310.12106},
    doi = {10.48550/arXiv.2310.12106},
    urldate = {2025-10-02},
    publisher = {arXiv},
    author = {Brown, Natalie C. and III, John Peter Campora and Granade, Cassandra and Heim, Bettina and Wernli, Stefan and Ryan-Anderson, Ciaran and Lucchetti, Dominic and Paetznick, Adam and Roetteler, Martin and Svore, Krysta and Chernoguzov, Alex},
    month = oct,
    year = {2023},
    note = {arXiv:2310.12106 [quant-ph]},
}

@article{campbell_magic-state_2012,
    title = {Magic-{State} {Distillation} in {All} {Prime} {Dimensions} {Using} {Quantum} {Reed}-{Muller} {Codes}},
    volume = {2},
    issn = {2160-3308},
    url = {https://link.aps.org/doi/10.1103/PhysRevX.2.041021},
    doi = {10.1103/PhysRevX.2.041021},
    number = {4},
    urldate = {2024-01-23},
    journal = {Physical Review X},
    author = {Campbell, Earl T. and Anwar, Hussain and Browne, Dan E.},
    month = dec,
    year = {2012},
    pages = {041021},
}

@article{aaronson_improved_2004,
    title = {Improved simulation of stabilizer circuits},
    volume = {70},
    copyright = {http://link.aps.org/licenses/aps-default-license},
    issn = {1050-2947, 1094-1622},
    url = {https://link.aps.org/doi/10.1103/PhysRevA.70.052328},
    doi = {10.1103/PhysRevA.70.052328},
    number = {5},
    urldate = {2025-10-14},
    journal = {Physical Review A},
    author = {Aaronson, Scott and Gottesman, Daniel},
    month = nov,
    year = {2004},
    pages = {052328},
}

@article{kubica_single-shot_2022,
    title = {Single-shot quantum error correction with the three-dimensional subsystem toric code},
    volume = {13},
    copyright = {2022 The Author(s)},
    issn = {2041-1723},
    url = {https://www.nature.com/articles/s41467-022-33923-4},
    doi = {10.1038/s41467-022-33923-4},
    number = {1},
    urldate = {2025-03-27},
    journal = {Nature Communications},
    author = {Kubica, Aleksander and Vasmer, Michael},
    month = oct,
    year = {2022},
    note = {Publisher: Nature Publishing Group},
    pages = {6272},
}

@article{delfosse_beyond_2022,
    title = {Beyond single-shot fault-tolerant quantum error correction},
    volume = {68},
    issn = {0018-9448, 1557-9654},
    url = {http://arxiv.org/abs/2002.05180},
    doi = {10.1109/TIT.2021.3120685},
    number = {1},
    urldate = {2025-03-07},
    journal = {IEEE Transactions on Information Theory},
    author = {Delfosse, Nicolas and Reichardt, Ben W. and Svore, Krysta M.},
    month = jan,
    year = {2022},
    note = {arXiv:2002.05180 [quant-ph]},
    pages = {287--301},
}

@misc{fazio_logical_2024,
    title = {Logical {Noise} {Bias} in {Magic} {State} {Injection}},
    url = {http://arxiv.org/abs/2401.10982},
    urldate = {2024-01-23},
    publisher = {arXiv},
    author = {Fazio, Nicholas and Harper, Robin and Bartlett, Stephen},
    month = jan,
    year = {2024},
    note = {arXiv:2401.10982 [quant-ph]},
}

@misc{ruiz_unfolded_2025,
    title = {Unfolded distillation: very low-cost magic state preparation for biased-noise qubits},
    shorttitle = {Unfolded distillation},
    url = {http://arxiv.org/abs/2507.12511},
    doi = {10.48550/arXiv.2507.12511},
    urldate = {2025-10-18},
    publisher = {arXiv},
    author = {Ruiz, Diego and Guillaud, Jérémie and Vuillot, Christophe and Mirrahimi, Mazyar},
    month = jul,
    year = {2025},
    note = {arXiv:2507.12511 [quant-ph]},
}

@misc{graham_mid-circuit_2023,
    title = {Mid-circuit measurements on a neutral atom quantum processor},
    url = {http://arxiv.org/abs/2303.10051},
    urldate = {2023-03-28},
    publisher = {arXiv},
    author = {Graham, T. M. and Phuttitarn, L. and Chinnarasu, R. and Song, Y. and Poole, C. and Jooya, K. and Scott, J. and Scott, A. and Eichler, P. and Saffman, M.},
    month = mar,
    year = {2023},
    note = {arXiv:2303.10051 [physics, physics:quant-ph]},
}

@misc{bluvstein_architectural_2025,
    title = {Architectural mechanisms of a universal fault-tolerant quantum computer},
    url = {http://arxiv.org/abs/2506.20661},
    doi = {10.48550/arXiv.2506.20661},
    urldate = {2025-10-18},
    publisher = {arXiv},
    author = {Bluvstein, Dolev and Geim, Alexandra A. and Li, Sophie H. and Evered, Simon J. and Ataides, J. Pablo Bonilla and Baranes, Gefen and Gu, Andi and Manovitz, Tom and Xu, Muqing and Kalinowski, Marcin and Majidy, Shayan and Kokail, Christian and Maskara, Nishad and Trapp, Elias C. and Stewart, Luke M. and Hollerith, Simon and Zhou, Hengyun and Gullans, Michael J. and Yelin, Susanne F. and Greiner, Markus and Vuletic, Vladan and Cain, Madelyn and Lukin, Mikhail D.},
    month = jun,
    year = {2025},
    note = {arXiv:2506.20661 [quant-ph]},
}
\end{document}